\documentclass[a4paper,11pt]{article}
\usepackage[latin1]{inputenc}
\usepackage{indentfirst}

\usepackage{amsfonts}
\usepackage{float}
\usepackage{latexsym,amssymb,amsmath}
\usepackage[dvips]{graphics,graphicx,epsfig,color}
\usepackage[lofdepth,lotdepth]{subfig}
\begin{document}
\thispagestyle{empty}
\title{\bf Seismic response in modern cities}
\author{Armand Wirgin\thanks{LMA, CNRS, UMR 7031, Aix-Marseille Univ, Centrale Marseille, F-13453 Marseille Cedex 13, France, ({\tt wirgin@lma.cnrs-mrs.fr})} }
\date{\today}
\maketitle
\begin{abstract}
  The proposed homogeneous flat-faced layer-like model of a city (termed overlayer), covering what is generally considered to be a dangerous site (from the point of view of seismic hazard) lends itself to an explicit theoretical analysis  of its  response to a seismic body wave radiated by distant sources. This study is carried out for: ground response of the complete site/overlayer configuration which is compared to the response of the configuration in which the overlayer is absent, response at the top of the layer for various  layer thicknesses, and  determination, as a function of frequency, of the fraction of incident flux that is dissipated in the overlayer, the underlying layer and, by radiation damping, in the hard half space. It is shown that  all of these entities are highly frequency-dependent and even large in certain frequency intervals, without any resonant (in the sense of mode excitation) phenomena coming into play. The results of this study also show that transfer functions do not necessarily reflect the global response in the built component of a city and that  more-appropriate energy-related functions, termed spectral absorptance (in the blocks of the city or their layer-like surrogate at the characteristic frequency of the seismic pulse) and absorptance (integral over frequency of the spectral absorptance), can increase with increasing city density  or increasing city height. In fact, it is shown that more than a third of the incident seismic energy can be sent into, and therefore cause serious damage to,  the built component. On the basis of these findings, it appears that the probable evolution of the morphology and constitutive properties of cities  with time will make the latter more vulnerable to damage and destruction when submitted to  seismic waves.
\end{abstract}
\newline
\newline
Abbreviated title: Seismic motion in  overlayers
\newline
\newline
Corresponding author: Armand Wirgin, \\ e-mail: wirgin@lma.cnrs-mrs.fr
\newpage
\tableofcontents
\newpage
\newpage
\section{Introduction}\label{intro}
Populations have an increasing tendency of residing in cities resulting in urban entities which become progressively denser  and whose buildings  become progressively taller. When a seismic wave hits such a city it will provoke a level of damage to the buildings therein which surely depends on the density of the city and the height of its buildings. Our  investigation is devoted to the evaluation of the effects of changes of  morphology (and the constitutive properties of the buildings) on seismic response in cities.
\subsection{Idealizations of buildings, blocks and their arrangement}
A city is, by definition, a rather large assembly of buildings. In  modern (and even some old) cities, the buildings are grouped into blocks separated, more or less periodically, by streets for the circulation of people and vehicles.

As is often the case in studies of the effects of earthquakes in cities \cite{wb96,ca01,gbc02,g05,gtw05,ks06,bg07,gw08,t10,tb11b,sn15,gc16} the buildings are homogenized (exceptions exist, as for instance in \cite{ll12}), which means that their constitutive properties, which vary greatly from one point to another within the buildings, are reduced to average (in some sense) constitutive parameters at all points of the structure. Moreover, the homogenized parameters and geometry vary from one building to another in a given block, so that it may be useful to further homogenize--this time the block-- by assigning an average building height and constitutive parameters to it.

The lateral spacing between blocks (or generic buildings) and the heights of these blocks (or generic buildings) vary from city to city and even from one area to another of a given city. It is thus of considerable interest  to determine how the  response of the city  to a seismic solicitation varies with these geometric parameters, notably the average city height  and the filling fraction  of the blocks relative to  that of the total city volume.
\subsection{Previous studies of seismic response in cities}\label{prevstud}
The problem of the seismic response in  cities   underlain by a soft layer (or soft basin, both being a common element of the sites of earthquake-prone cities) covering a very hard half space underground, has been treated in a surprisingly-modest  number of publications, considering the social and economic importance of the subject. The increasing tendency is to resort to numerical  (separation of variables \cite{wb96,g05,gtw05,gw08}, boundary element \cite{ca01,k04,ks06},  finite-difference or finite element \cite{t10,g05,gtw05,y10,tb11b,ll12}, spectral element \cite{kj06,gc16}, substructuring \cite{ca01,g05,ll12,y10}, etc.) methods of analysis. However, it is difficult to obtain simple, physical interpretations of the observed/computed phenomena from parametric studies  based on purely-numerical procedures involving variation of scores of parameters. This is all the more true than the presence of the city over the site gives rise to complex physical effects (notably coupling to structural modes or to what may appear as modes)  that translate to  mathematical (and consequently, theoretical) complexity \cite{g05,gw08}. Consequently, the still largely-unanswered question is: what are the main factors that condition the elastodynamic wavefield in the built components,  and therefore the amount and distribution of damage, in the city.

 The soft layer component of what lies beneath the ground, is certainly a factor which aggravates the effects of earthquakes in cities. It  is a very complex medium that most of the previously-cited publications  also homogenize,  partly because  one knows relatively-little of its composition and geometry.  Fortunately, more is known about the above-ground structure (i.e., the city)  because it is visible  and/or records have been made of its composition and geometry. Thus,  one should strive to incorporate this knowledge, in as simple and efficient manner  as possible, in any attempt to account for the  seismic response in the city.

 A drastic simplification is to replace the ground (on which the city rests) or  the wavy stress-free outer boundary of the city by a flat surface on which an impedance boundary condition prevails. Naturally, the main problem is how to relate this impedance to the geometrical and constitutive properties of the buildings/blocks. Boutin and colleagues \cite{br04,r06,sb16} have succeeded in doing this by the employment of a homogenization technique which   is thought to be valid for lateral dimensions (building width and separations that are small compared to the smallest wavelength in the spectrum of the seismic pulse. Other simplified models and references are described, and can be found, in  \cite{w16}.
\subsection{Preview of what we want to accomplish and how to do this}\label{prevstud}
 By a quite different approach (from that of Boutin et al.) based on an essentially low-frequency, high block width over period ratio approximation procedure, one can demonstrate theoretically  that a city, consisting of  a periodic distribution of identical buildings or blocks, responds to  a seismic disturbance  in much the same manner as a homogeneous layer whose  thickness and constitutive properties  are  simply-related to the geometric and physical properties of the original city.   The predictions of city response via the layer model are, for dense cities at low solicitation frequencies, similar to those of Boutin and colleagues.  and, as shown herein, to the computed responses obtained from  rigorous periodic city models \cite{g05,gw08}.

Using both the periodic block and layer models, which both automatically account for Site-(above ground) Structure Interaction (SSI)\cite{mg00}, we shall carry out  computations similar to those of Kham et al. \cite{ks06} in order to:\\
(a) find out how changes in city density (by adding more (identical) buildings in a given area on the ground) influence the seismic response at the bottom (i.e., ground level) and top (roof level of the buildings/blocks) of the city\\
(b) find out whether SSI in cities \cite{h57,it14} can be qualified as a beneficial \cite{mg00,ks06,bg07} or detrimental \cite{mg00} effect,\\
(c) determine whether the examination of measurement entities such as ground motion  can give a decisive answer to question  (b),\\
(d) find out how much energy is injected into a city during an earthquake and thus obtain an idea of the resultant damage in the city.
\section{Basic features of the  periodic block model}
In a cartesian coordinate system $Oxyz$, with origin at $O$, the ground, assumed to be flat, occupies the entire plane $z=0$. In the absence of the city, the half-space above the ground, filled with air, is assumed to be occupied by the vacuum. In the presence of the city, the outer boundaries of the blocks are in contact with the air (replaced by the vacumn).

The earthquake sources are assumed to be located in the half-space below and infinitely-distant from the ground so that the seismic (pulse-like) solicitation takes the form of a body (plane) wave in the neighborhood of the ground. This plane wavefield is assumed to be of the $SH$ variety, so that only one (i.e., the $y$-) component of the incident displacement field is non-nil, i.e., $\mathbf{u}^{i}=(0,u^{i}(\mathbf{x},\omega),0)$, wherein $\mathbf{x}=(x,z)$ and $\omega=2\pi f$  the angular frequency, $f$ the frequency. The solicitation is characterized by three parameters (angle of incidence, and two parameters characterizing the pulse).
\begin{figure}[ht]
\begin{center}
\includegraphics[width=0.55\textwidth]{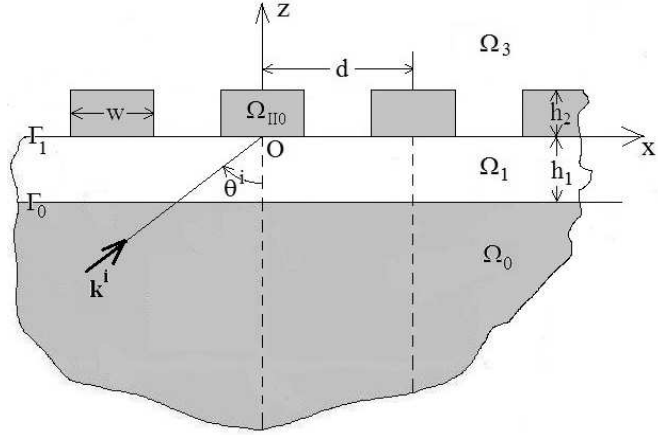}
\caption{Sagittal plane view of  configuration $\mathcal{C}_{II}$ comprising a rather idealized city above (site) configuration $\mathcal{C}_{1}$, whose seismic response is of concern herein. The site is composed of a flat-faced (usually soft) solid layer (white in the figures) underlain by a very hard solid half space (dark grey in the figures). The ground is at $z=0$. This above-ground structure  gives rise, under certain circumstances,  to a seismic response similar to that of a city in the form of  a flat-faced, homogeneous layer, depicted in fig. \ref{sitewithlayerlikecity}.}
\label{sitewithblocklikecity}
\end{center}
\end{figure}

 The  configuration termed $\mathcal{C}_{II}$, depicted in fig. \ref{sitewithblocklikecity}, consists of a city above a soft, homogeneous (soil) flat-faced layer  overlying, and in welded contact with, a hard half-space.  The upper face of the soil layer is the ground located at $z=0$ and its planar, lower face is described by $z=-h_{1}$, with $h_{1}$ its thickness. This  is   typical (although ideally-so) of the sites on which the majority of earthquake-prone  cities are built. Our  city is assumed to be composed of a periodic (along $x$, with period $d$) set of identical (in geometry and composition), mutually-parallel blocks  that are infinitely long along $y$. These structures are assumed to be in welded contact with the soil layer across the ground.

The most general problem considered herein is that of determining the seismic response in $\mathcal{C}_{II}$  This  involves more than a doezn configurational parameters plus three solicitational parameters, which means that,  if one strives to obtain simple physical descriptions of computed response via parametric studies involving variation of these  parameters, he will be faced with a very confusing task. A merit of the study of Kham et al. \cite{ks06} is to show that important features of the seismic response in cities have to do with the single parameter of the city which is its density (i.e., the ratio of the area occupied by the buildings to the total area of the city). One can also guess that another important parameter is the average height of the buildings.

Since neither the incident wavefield nor the geometric features of the site  depend on $y$, the total wavefield $\mathbf{u}$ depends only on $x$ and $z$, which means that the to-be-considered problem is 2D (although the terrestrial model is 1D) and can be examined in the sagittal $x-z$ plane. Fig. \ref{sitewithblocklikecity} depicts the problem involving $\mathcal{C}_{II}$ in this sagittal plane in which: $\Omega_{0}$ is the half-space domain occupied by the hard, homogeneous, isotropic material $\mathcal{M}^{[0]}$,  $\Omega_{1}$  the layer-like domain  occupied by the soft, linear, homogeneous, isotropic material $\mathcal{M}^{[1]}$, $\Omega_{II}=\cup_{n\in\mathbb{Z}}\Omega_{IIn}$  the domain occupied by the city, $\Omega_{IIn}$ the domain of the $n$-th  block of rectangular cross section (width $w$ and height $h_{2}$) occupied by a  relatively-soft, linear, homogeneous, isotropic material $\mathcal{M}^{[2]}$ (note that this is a considerable idealization since a block is composed of a number of buildings (in \cite{ks06}, and in the numerical examples given further on relative to this publication, this number is one) and the materials of these buildings   are neither linear, nor isotropic,  nor homogeneous), and $\Omega_{3}$ the  remaining portion  of $\mathbb{R}^{2}$ occupied by the vacuum. In the sagittal plane, the interface (i.e., the line $z=-h_{2}$) between $\Omega_{0}$ and $\Omega_{1}$ is designated by $\Gamma_{0}$, the ground (i.e., the line $z=0$) by $\Gamma_{1}$, and the interface  between  $\Omega_{2}$ and $\Omega_{3}$  by $\Gamma_{II}=\cup_{n\in\mathbb{Z}}\Gamma_{IIn}$, in which $\Gamma_{II0}$ is the portion of $\Gamma_{II}$ included between $x=-d/2$ and $x=d/2$ (see fig. \ref{sitewithblocklikecity} for the sagittal plane view of $\mathcal{C}_{II}$).

The three media are assumed to be  non-dispersive over the range of frequencies  of interest. The shear moduli $\mu^{[l]}$ of $\mathcal{M}^{[l]}~;l=0,1,2$ are assumed to be real. The shear-wave velocities $\beta^{[l]}~;~l=1,2$ are complex, i.e., $\beta^{[l]}=\beta^{'[l]}+i\beta^{''[l]}$, with $\beta^{'[l]}\ge 0$, $\beta^{''[l]}\le 0$, $\beta^{[l]}=\sqrt{\frac{\mu^{[l]}}{\rho^{[l]}}}$, and  $\rho^{[l]}$ the mass density. The shear-wave velocity $\beta^{[0]}$ is real, i.e., $\beta^{''[0]}=0$.

The  wavevector $\mathbf{k}^{i}$ of the plane wave solicitation lies in the  sagittal plane and is of the form $\mathbf{k}^{i}=(k_{x}^{i},k_{z}^{i})=\left(k^{[0]}s^{i},k^{[0]}c^{i}\right)$
wherein  $\theta^{i}$ is the angle of incidence (see fig. \ref{sitewithblocklikecity}), $s^{i}=\sin\theta^{i}$, $c^{i}=\cos\theta^{i}$ and $k^{[l]}=\omega/\beta^{[l]}$.

The rigorous theory of the seismic response of $\mathcal{C}_{II}$ was given in detail in \cite{g05,gw08}.
\section{Basic features of the  layer model(s)}
Actually, we shall consider three types of layer configurations, all of which lend themselves to a simple, although rigorous, analysis of their seismic response. The reason for considering three configurations is that, traditionally, to appreciate the specific influence of the presence of a city above a given site, its response is compared to that of the site in the absence of the city. But due to the usual presence of the soft basin or layer below the ground, even the seismic response of such a site is complex so that the absolute reference is taken to be that of a site consisting simply of a hard half space below the ground.
\begin{figure}[ht]
\begin{center}
\includegraphics[width=0.65\textwidth]{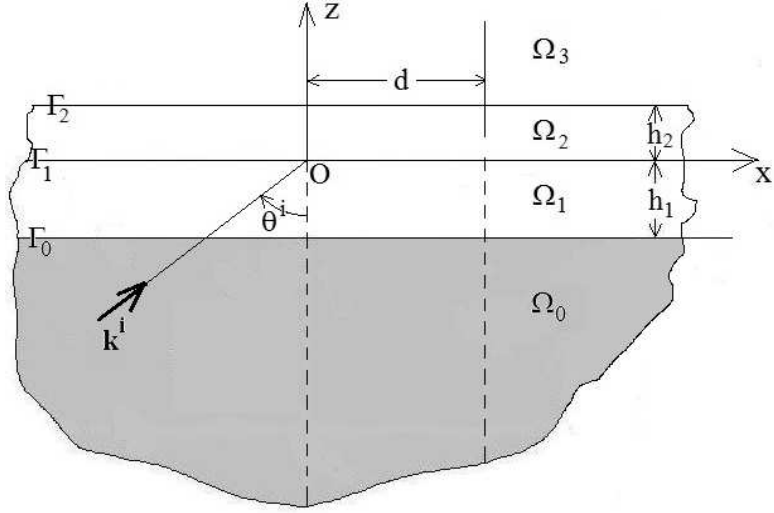}
\caption{Sagittal plane view of the layer configuration $\mathcal{C}_{2}$ of the city comprising a flat-faced homogeneous layer (i.e., the overlayer which simulates the presence of the city)  above (site) configuration $\mathcal{C}_{1}$. This site (i.e., $\mathcal{C}_{1}$) is composed of a flat-faced (usually soft) solid layer (white in the figures) underlain by a very hard solid half space (dark grey in the figures).}
\label{sitewithlayerlikecity}
\end{center}
\end{figure}

The doulble-layer configuration, termed  $\mathcal{C}_{2}$ (see fig. \ref{sitewithlayerlikecity}), whose seismic response is thought (and shown further on) to be similar to that of the  periodic block model of the city $C_{II}$, consists of a homogeneous layer (termed overlayer, and which replaces the periodic-block city) above (and in welded contact) with the same  site \cite{st28} as that of $C_{II}$, i.e., the configuration $C_{II}$ in the absence of the blocks. The thickness of the overlayer equals the height $h_{2}$ of the blocks in $C_{II}$ and the shear body wave velocity $b^{[2]}$ in the overlayer equals  the shear body wave velocity  $\beta^{[2]}$ in the  the generic block of the $C_{II}$ city, whereas the shear modulus $m^{[2]}$ of $\mathcal{C}_{2}$ is simply
\begin{equation}\label{0-000}
  m^{[2]}=\mu^{[2]}\phi~,
\end{equation}
with $\phi$ the filling factor
\begin{equation}\label{0-002}
    \phi=\frac{w}{d}~,
\end{equation}
and $w$ and $d$  the previously-defined geometric parameters of the generic block of $C_{II}$, whereas $\mu^{[2]}$ is the shear modulus in this block. The parameters of the site (termed $C_{1}$) in $C_{2}$ are the same as those in $C_{II}$, i.e., $b^{[j]}=\beta^{[j]}$, $m^{[j]}=\mu^{[j]}~;~j=0,1$ and $h_{1}$ designating as before the thickness of the underlayer, with the understanding that the ground is located at $z=0$ in both configurations.  Finally, the solicitation of $C_{2}$ is identical to that of $C_{II}$.

The  second configuration, $C_{1}$, is $C_{2}$ without the overlayer, or $C_{2}$ in the limit $h_{2}\rightarrow 0$. The third configuration, $C_{0}$, is $C_{1}$ without the underlayer, or $C_{1}$ in the limit $h_{1}\rightarrow 0$. In both of these configurations, the ground is located at $z=0$ and the solicitation is as in $C_{2}$.

We assume herein that
\begin{equation}\label{0-001}
|b^{''[l]}/b^{'[l]}|<<1~;~l=1,2~.
\end{equation}
Condition (\ref{0-001}) is invoked only for the sake of simplifying certain aspects of the subsequent analysis, but is by no means essential for most of what follows.

The soft nature of the the two layers signifies that:
\begin{equation}\label{0-003}
0<m^{[j]}b^{[0]}/m^{[0]}b^{'[j]}<1~;~j=1,2~.
\end{equation}
The total wavefield $u(\mathbf{x};\omega)$ in $\Omega^{[l]}$ is designated by $u^{[l]}(\mathbf{x\mathbf{}};\omega)$. The incident wavefield is
\begin{equation}\label{1-000}
u^{i}(\mathbf{x},\omega)=A^{[0]+}(\omega)\exp[i(k_{x}x+k_{z}^{[0]}z)]~,
\end{equation}
wherein $A^{[0]+}(\omega)$ is the spectral amplitude of the solicitation and $\omega=2\pi f$, with $f$ the frequency.

Using the techniques of  \cite{st28,ejp57,k64}, one finds that the solution of the boundary value problem connected with the above description of the physical problem is:
\begin{equation}\label{1-010}
u_{m}^{[l]}(\mathbf{x},\omega)=u_{m}^{[l]+}(\mathbf{x},\omega)+u_{m}^{[l]-}(\mathbf{x},\omega)~;~l=0,1,2~;~m=0,1,2~,
\end{equation}
in which the superscripts $+$ and $-$ refer to  upgoing and downgoing  waves respectively), $u_{m}^{[l]}(\mathbf{x};\omega)$ is the total displacement field in $\Omega_{l}$ for configuration $\mathcal{C}_{m}$, and $u_{m}^{[0]+}=u^{i}~;~m=0,1,2$, with
\begin{equation}\label{1-015}
u_{m}^{[l]\pm}(\mathbf{x},\omega)=A_{m}^{[l]\pm}(\omega)\exp[i(k_{x}x\pm k_{z}^{[l]}z)]~,
\end{equation}
 wherein
\begin{equation}\label{1-25}
k_{x}=k_{x}^{i},~~,~~k_{z}^{[l]}=\sqrt{(k^{[l]})^{2}-(k_{x})^{2}}~~;~~ \Re k_{z}^{[l]}\ge0~,~\Im k_{z}^{[l]}\ge0 \text{ ~for~} \omega\ge 0~.
\end{equation}
The temporal response $U_{m}^{[l]}(\mathbf{x};t)$ is obtained from the spectral response $u_{m}^{[l]}(\mathbf{x};\omega)$ via
\begin{equation}\label{1-070}
U_{m}^{[l]}(\mathbf{x};t)=2\Re\int_{0}^{\infty}u_{m}^{[l]}(\mathbf{x};\omega)\exp(-i\omega t)d\omega~.
\end{equation}
The spectrum of the (Ricker) pulse-like solicitation is of the form
\begin{equation}\label{1-080}
A^{[0]+}(\omega)=\Big(\frac{f}{\nu}\Big)^{2}\exp\Big[i2\pi f\tau-\Big(\Big(\frac{f}{\nu}\Big)^{2}-1\Big)\Big]
~,
\end{equation}
wherein $\nu$ is the frequency at which the pulse is at its maximum (termed hereafter 'characteristic frequency'), and  $\tau$ is related to the onset time of the pulse.

It is easily shown \cite{ejp57,k64} that:
\begin{equation}\label{1-110}
A_{2}^{[2]-}=A_{2}^{[2]+}(e^{[22]})^{2}~~,~~A_{2}^{[2]+}=A^{[0]+}\big(e^{[01]}e^{[22]}\big)^{-1}(D_{2})^{-1}~.
\end{equation}
\begin{equation}\label{1-125}
A_{2}^{[1]\pm}=
A^{[0]+}\big(e^{[01}\big)^{-1}
\big(C^{[22]}\mp i S^{[22]}g^{[21]}\big)
(D_{2})^{-1}~,
\end{equation}
\begin{equation}\label{1-130}
A_{2}^{[0]-}=A^{[0]+}\big(e^{[01]}\big)^{-2}\big[C^{[22]}\big(C^{[11]}+g^{[10]}iS^{[11]}\big)+
g^{[21]}iS^{[22]}\big(g^{[10]}C^{[11]}+iS^{[11]}\big)\big](D_{2})^{-1}
~,
\end{equation}
wherein:
\begin{equation}\label{1-140}
D_{2}=C^{[22]}\big(C^{[11]}-g^{[10]}iS^{[11]}\big)-g^{[21]}iS^{[22]}\big(g^{[10]}C^{[11]}-iS^{[11]}\big)~.
\end{equation}
 and $e^{[jk]}=\exp(ik_{z}^{[j]}h_{k})$, ~ $C^{[jk]}=\cos (k_{z}^{[j]}h_{k})$,~ $S^{[jk]}=\sin (k_{z}^{[j]}h_{k})$,~  $g^{[jk]}=m^{[j]}k^{[j]}/m^{[k]}k^{[k]}=$ $m^{[j]}b^{[k]}/m^{[k]}b^{[j]}$.

 It then follows (for $\mathcal{C}_{1}$) by taking $h_{2}\rightarrow 0$:
\begin{equation}\label{1-150}
A_{1}^{[1]-}=A_{1}^{[1]+}~~,~~A_{1}^{[1]+}=A^{[0]+}\big(e^{[01]}\big)^{-1}D_{1}^{-1}~.
\end{equation}
\begin{equation}\label{1-170}
A_{1}^{[0]-}=A^{[0]+}\big(e^{[01]}\big)^{-2}\big[C^{[11]}+g^{[10]}iS^{[11]}\big]D_{1}^{-1}~~,~~D_{1}=C^{[11]}-g^{[10]}iS^{[11]}~.
\end{equation}

Finally, taking additionally $h_{1}\rightarrow 0$ (for $\mathcal{C}_{0}$) results in:
\begin{equation}\label{1-190}
A_{0}^{[0]-}=A^{[0]+}~.
\end{equation}
In all the numerical examples of this paper we shall adopt (unless stated otherwise)  the geometric, constitutive and solicitational parameters of our study \cite{w16} which  originate in those of the publication of Kham et al. \cite{ks06}: $b^{[0]}=\beta^{[0]}=1000 ~ms^{-1}$, $m^{[0]}=\mu^{[0]}=2\times 10^{9} Pa$, $b^{[1]'}=\beta^{[1]'}=200~ms^{-1}$, $b^{[1]''}=\beta^{[1]''}=-4~ms^{-1}$, $m^{[1]}=\mu^{[1]}=7.2\times 10^{7} Pa$, $h_{1}=25~m$, $b^{2]'}=\beta^{2]'}=240~ms^{-1}$, $b^{[2]''}=\beta^{[2]''}=-12~ms^{-1}$, $\mu^{[2]}=1.44\times 10^{7} Pa$, $\theta^{i}=0^{\circ}$, $\nu=2~Hz$, and $\tau=1~s$.
\section{Interrelations of the ground displacements in configurations $\mathcal{C}_{0}$, $\mathcal{C}_{1}$ and $\mathcal{C}_{2}$}
As underlined previously, it is customary to give a measure of the seismic response in a city by the way the presence of the  city modifies the ground displacement (or velocity or acceleration). Similarly, one measures the effect of the underlayer by comparing the ground displacement of $C_{1}$ to that of $C_{0}$.
\subsection{Field on the ground in configuration $\mathcal{C}_{0}$}
Eqs. (\ref{1-010}), (\ref{1-015}) and (\ref{1-190}) yield
\begin{equation}\label{2-015}
\Big\|u_{0}^{[0]}(x,0;\omega)\Big\|=2\|A_{0}^{[0]+}(\omega)\|~.
\end{equation}
This entity is often considered to be the reference by which transfer functions are defined because it is assumed that it can be measured and does not depend on disturbing features such as uneven ground, heterogeneity of the medium that lies below the ground, uncertainty concerning the seismic solicitation. In practice, this may not be the case because the underground is neither homogeneous nor infinitely hard in the vicinity of where the measurement is made,  the ground is never really flat in the neighborhood of the seismometer and  objects located thereon (such as buildings) which perturb the field, and the seismic sources are never infinitely far from (and beneath) the ground.

For an arbitrary site giving rise to a measured field $u(\mathbf{x},\omega)$, the usual definition of the modulus of the transfer function  is  $\Big\|u(\mathbf{x},\omega)\Big\|/\Big\|u_{0}^{[0]}(x,0,\omega)\Big\|
=\Big\|u(\mathbf{x},\omega)\Big\|/2\Big\|A_{0}^{[0]+}(\omega)\Big\|$. Henceforth, we suppress the factor 2 so as to define the modulus of transfer function $\mathcal{T}$ at a given location $\mathbf{x}$ as $\mathcal{T}=\Big\|u(\mathbf{x},\omega)\Big\|/\Big|\|A_{0}^{[0]+}(\omega)\Big\|$ wherein $\|A_{0}^{[0]+}(\omega)\|$ is obtained from (the measurement of $\Big\|u_{0}^{[0]}(x,0,\omega)\Big\|$.

Consequently, the modulus of the transfer function on  flat ground over a hard half space is
\begin{equation}\label{2-020}
\mathcal{T}_{0}(x,0,\omega)=\Big\|\frac{u_{0}^{[0]}(x,0;\omega)}{A^{[0]+}(\omega)}\Big\|=2~.
\end{equation}
Note that this result is independent of the angular frequency $\omega$ and $x$.
\subsection{Field on the ground in configuration $\mathcal{C}_{1}$}
Eqs. (\ref{1-010}), (\ref{1-015}) and (\ref{1-150}) yield,  under the assumption  (\ref{0-001}) (amounting here to considering $b^{[1]}$, $k_{z}^{[1]}$ and $g^{[10]}$ to be real)
\begin{equation}\label{2-050}
\mathcal{T}_{1}(x,0,\omega)=\Big\|\frac{u_{1}^{[1]}(x,0;\omega)}{A^{[0]+}(\omega)}\Big\|=\Big\|\frac{2A_{1}^{[1]+}}{A^{[0]+}}\Big\|\approx
\frac{2}{\sqrt{(C^{[11]})^{2}+(g^{[10]}S^{[11]})^{2}}}
~.
\end{equation}
However, the assumption (\ref{0-003}) implies that $0<g^{[10]}<1)$ whence
from  (\ref{2-050}) and (\ref{2-020})
\begin{equation}\label{2-060}
\mathcal{T}_{1}(x,0,\omega)\ge \mathcal{T}_{0}(x,0,\omega)~,
\end{equation}
or, in other words: {\it the modulus of the transfer function on the ground of configuration $\mathcal{C}_{1}$ is greater than or equal to the modulus of the transfer function on the ground of configuration $\mathcal{C}_{0}$, this being so at all frequencies}. This amplification of ground motion \cite{lc06} is why a site such as $\mathcal{C}_{1}$ is qualified as being 'dangerous', i.e., for a massless observer located on the ground compared to what he would sense if the medium underneath the ground were uniformly hard.

$\|u_{1}^{[1]}(x,0;\omega)\|$ is a quasi-periodic function of $\omega$ that attains its maximal values when  $\|S^{[11]}\|=\pm 1$, i.e. for  frequencies
\begin{equation}\label{2-070}
f_{1n}=\frac{n b^{[0]}}{4h_{1}\eta^{[1]}}~~;~~n=1,3,5,...~.
\end{equation}
wherein $\eta^{[1]}=\sqrt{\big(\frac{b^{[0]}}{b^{'[1]}}\big)^{2}-\big(s^{i}\big)^{2}}$.
By convention, the so-called soil frequency  of the site in configuration $\mathcal{C}_{1}$ is obtained for $\theta^{i}=0~\Rightarrow~s^{i}=0$ and $n=1$ and is given by
\begin{equation}\label{2-080}
f_{11}=\frac{b^{'[1]}}{4h_{1}}~.
\end{equation}
Note that  the modulus of the transfer function on $z=0$ at these  frequencies  is
\begin{equation}\label{2-085}
\mathcal{T}_{1}(x,0,\omega_{1n})=\Big\|\frac{u_{1}^{[1]}(x,0;\omega_{1n})}{A^{[0]+}(\omega_{n})}\Big\|\approx\Big\|\frac{2}{g^{[10]}}\Big\|~,
\end{equation}
(wherein $\omega_{1n}=2\pi f_{1n}$) which is independent of $h_{1}$.

\begin{figure}[ht]
\begin{center}
\includegraphics[width=0.65\textwidth]{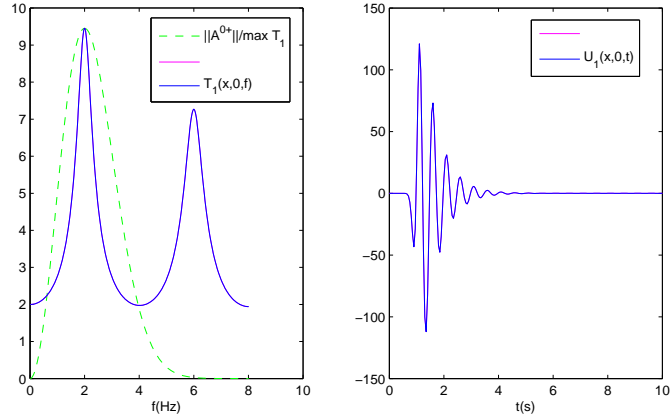}
\caption{Motion on the ground for the configuration $\mathcal{C}_{1}$ of a single soft layer over a hard half space. Left panel: the blue curve is the modulus of the transfer function, and the dashed green curve denotes the modulus of the Ricker spectrum of the incident pulse. Right panel: the time domain response to the Ricker pulse.}
\label{f010}
\end{center}
\end{figure}
In the left-hand panel of fig. \ref{f010}, we have  depicted $\|A^{[0]+}(\omega)\|/\max\|u_{1}^{[1]}(x,0;\omega)\|$, i.e., the normalized spectrum of the seismic pulse, assuming, as in \cite{ks06}, that this pulse is Ricker-like and
$\nu$ (i.e.,  the frequency at which the pulse is at its maximum) is $\nu=f_{11}=2 Hz$)(i.e., the same choice as in \cite{ks06}). Using this information, the time domain displacement on the ground is computed via (\ref{1-070}), this function being depicted in the right-hand panel of fig. \ref{f010}.

A word is here in order about the peaks in fig. \ref{f010}. The response in the neighborhoods of $f_{1n}$ is often qualified \cite{bb80,g82,ks06,bg07,sb16} as being "resonant" and the $f_{1n}$ in (\ref{2-070}) to be "eigenfrequencies" or "natural frequencies". This would be true if a mode (such as a Love or Rayleigh eigenmode) of the structure were being excited at these frequencies, in which case the response would be infinite at the resonance frequencies in the absence of dissipation in the two media of the configuration. But this is clearly not true because $D_{1}(\omega)$ (which is the denomùinator of the expression for displacement response) does not vanish at any frequency, even when the constitutive parameters of the two media are real. Thus, the response peaks are not the result of a  "resonance" (i.e., coupling to a mode) but merely the trace (visible in fig. \ref{f010}) of constructive {\it interference} of the two plane waves within the layer. The excitation of a modal resonance requires either a more complex solicitation,  such as the wave radiated by a source not too far from the flat layer/hard half space interface \cite{gw05a,gw05b} or  a rough \cite{w88} or non-planar interface \cite{w95} and/or a rough or modulated-impedance ground \cite{wg06a,wg06b}. The nature of the resonance peaks in a periodic block model of a city subjected to an earthquake has been explained in detail in \cite{gw08}.
\subsection{Field on the ground in configuration $\mathcal{C}_{2}$}
The modulus of the transfer function of ground  motion is
\begin{equation}\label{2-120}
\mathcal{T}_{2}(x,0,\omega)=\Big\|\frac{u_{2}^{[2]}(x,0,\omega)}{A^{[0]+}(\omega)}\Big\|=2\Big\|\frac{A_{2}^{[2]+}(\omega)e^{[22]}(\omega)C^{[22]}(\omega)}{A^{[0]+}(\omega)}\Big\|~,
\end{equation}
Again consider the  situation in which all the media are nearly lossless. Then, at the one-layer frequencies $\omega_{1n}=2\pi f_{1n}$ defined in (\ref{2-070}), at which $S^{11}=\pm 1$ and $C^{11}=0$:
\begin{equation}\label{2-130}
\mathcal{T}_{2}(x,0,\omega_{1n})=\Big\|\frac{u_{2}^{[2]}(x,0,\omega_{1n})}{A^{[0]+}(\omega_{1n})}\Big\|\approx\Big\|\frac{2}{g^{[10]}}\Big\|
\sqrt{\frac{\big(C^{[22]}(\omega_{1n})\big)^{2}}{\big(g^{[21]}g^{[01]}\big)^{2}\big(S^{[22]}(\omega_{1n})\big)^{2}+
\big(C^{[22]}(\omega_{1n})\big)^{2}}}\le\Big\|\frac{2}{g^{[10]}}\Big\|~,
\end{equation}
or, recalling (\ref{2-085}):
\begin{equation}\label{2-150}
\mathcal{T}_{2}(x,0,\omega_{1n})\le \mathcal{T}_{1}(x,0,\omega_{1n})
~,
\end{equation}
wherein $ \mathcal{T}_{1}(x,0,\omega_{1n})$ is the modulus of the transfer function on the ground of the one-layer configuration (i.e., the one in which the overlayer is absent) at the one-layer frequencies $\omega_{1n}$. This result is also obtained in the case of slightly-lossy layers and means that {\it the presence of a homogeneous layer on top of a given one-layer/half space site {\it reduces the ground motion} at the soil frequency of this site}, this effect having also been  discovered {\it numerically} for an overlayer in the form of a city composed of a row of identical blocks in \cite{ks06,bg07}.

Note that the  transfer function on $z=0$ at the soil frequency is now dependent on $h_{1}$ and also on $h_{2}$. Note also that the effect has been shown to occur  only at the soil frequency of the site, which means that it is possible (as will be shown hereafter) for the ground motion to be larger in the presence of the overlayer than in its absence at some frequencies other than the soil frequency of the site.

We now demonstrate the existence of the  effect on a numerical example. The overlayer (simulating a city) is chosen to have the constitutive  properties of a building in  \cite{ks06}. The city  properties are  different from those of the soil, as is likely to occur in reality.
\begin{figure}[ht]
\begin{center}
\includegraphics[width=0.65\textwidth]{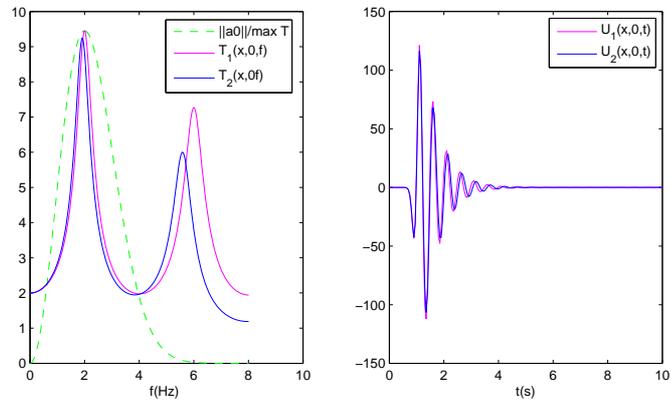}
\caption{Motion  on the ground for the configuration $\mathcal{C}_{2}$ of  two soft layers over a hard half space, the composition of the two layers being different. Left panel: the magenta curve is the modulus of the transfer function in the absence of the overlayer, and the blue curve is the modulus of the transfer function in the presence of the uppermost layer, whereas the dashed green curve denotes the modulus of the Ricker spectrum of the incident pulse. Right panel: the time domain responses (same meaning for the colors) to the Ricker pulse.  $h_{2}=7.5~m$}
\label{f090}
  \end{center}
\end{figure}
Fig. \ref{f090} tells us what happens at ground level  when we put a homogeneous overlayer on the ground. We   observe  that the two peaks of response on ground level apparently shift to lower frequency and diminish in amplitude at the site soil frequencies $f_{1n}$. An attempt to explain the  apparent frequency shift will be made in sect. \ref{freqsh}.
\clearpage
\newpage
\section{Spectral features of the responses on the ground  and top of the overlayer in configuration $\mathcal{C}_{2}$}
%
\subsection{Top versus bottom response in the overlayer}
It is found, assuming as is usual that the media are lossless, that
\begin{equation}\label{2-170}
\|u_{2}^{[2]}(x,h_{2},\omega)\|=\|2A^{[2]+}(\omega)\|~~,~~\|u_{2}^{[2]}(x,0,\omega)\|=\|2A^{[2]+}(\omega)C^{[22]}\|~,~,
\end{equation}
which, because $\|C^{[22]}\|\le 1$,  shows that {\it in configuration $\mathcal{C}_{2}$, the modulus of the field on the top of the overlayer is generally larger than the modulus of the field on the ground, this being so whatever the frequency}.
\subsection{A possible explanation of the apparent frequency shifts and amplitude reduction of maxima of response in $\mathcal{C}_{2}$}\label{freqsh}
 The transfer function of the field in the overlayer is largely conditioned by $D_{2}^{-1}$, which means that $\mathcal{T}_{2}$  is all the larger the smaller is $\|D_{2}\|$.

$D_{2}$ can be re-written as:
\begin{equation}\label{5-045}
D_{2}=\Big[\cos(k_{z}^{[2]}h_{2}+k_{z}^{[1]}h_{1}))-ig^{[10]}\sin(k_{z}^{[2]}h_{2}+k_{z}^{[1]}h_{1})\Big]+
(1-g^{[21]})S^{[22]}\big(S^{[11]}+ig^{[10]}C^{[11]}\big)~,
\end{equation}
The modulus of the term $[~]$  is minimal when $\Re\cos(k_{z}^{[2]}h_{2}+k_{z}^{[1]}h_{1}))\approx 0$ which occurs for
 frequencies
\begin{equation}\label{5-060}
f_{2l}=\frac{l\beta^{[0]}}{4\big(\eta^{[2]}h_{2}+\eta^{[1]}h_{1}\big)}~;~l=1,3,5,...~.
\end{equation}
Thus, when $1-g^{[21]}=0$, $\|D_{2}\|$ will be minimal, and the modulus of the field in the overlayer (including the ground and top of the overlayer)  will be maximal, at frequencies $f_{2l}$. These frequencies, contrary to the so-called characteristic frequency of the buildings \cite{09gmv}, depend not only on the height and velocity in the buildings, but also on the underlayer characteristics, in conformity with the idea of Soil-(above ground) Structure Interaction (SSI)\cite{mg00}.

We saw previously  that the the modulus of the field on the ground in the absence of the overlayer is maximal for frequencies $f_{1l}=\frac{l\beta^{[0]}}{\eta^{[1]}h_{1}}~;~l=1,3,5,...$, and since
\begin{equation}\label{5-070}
\frac{1}{\eta^{[2]}h_{2}+\eta^{[1]}h_{1}}\le \frac{1}{\eta^{[1]}h_{1}}~,
\end{equation}
it ensues that
\begin{equation}\label{5-080}
f_{2l}\le f_{1l}~;~l=1,3,5,...~,
\end{equation}
\begin{figure}[ht]
\begin{center}
\includegraphics[width=11cm] {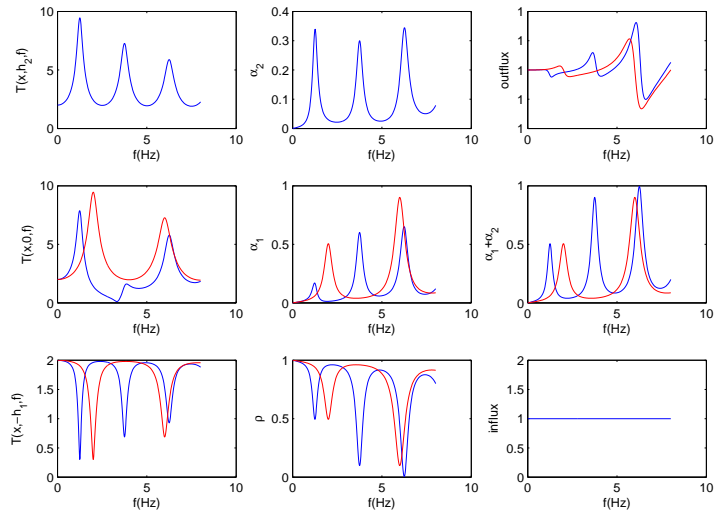}
 \caption{Frequency domain response functions for the double-layer configuration $\mathcal{C}_{2}$ (blue) and the single-layer configuration $\mathcal{C}_{1}$ (red). The (1,1) panel depicts the modulus of the top displacement transfer function, the (2,1) panel depicts the modulus of the ground displacement transfer function, the (3,1) panel depicts the modulus of the displacement transfer function on the interface between the soft underlayer and the half-space. The other panels, which involve entities that are not pertinent to the present discussion (but which are defined later on), concern the following. The (1,2) panel depicts the spectral absorptance $\alpha_{2}$ in the overlayer, the (2,2) panel depicts the spectral absorptance $\alpha_{1}$ in the underlayer and the (3,2) panel depicts the spectral reflectance $\rho$ in the halfspace. The (1,3) panel depicts the output flux, the (2,3) panel depicts the total spectral absorptance $\alpha_{1}+\alpha_{2}$, and the (3,3) panel depicts the input flux. $h_{2}=15~m$. Note (graphically) that the $l=1$ maximum of top response is at $1.245~Hz$ and it amplitude is $9.453~a.u.$, the $l=3$ maximum of top response is at $3.749~Hz$ and its amplitude is $7.269~a.u.$.}
  \label{triv2}
  \end{center}
\end{figure}
which means (notably because, as was assumed at the outset that $b^{[j]}$, and therefore $\eta^{[j]}$, do not depend on frequency) that the maxima of response of the configuration in the presence of  the overlayer occur at generally lower frequencies than those (for the same values of the index $l$) of the configuration without the overlayer, this being strictly-true only  in the case  $1-g^{[21]}=0$.

\begin{figure}[ptb]
\begin{center}
\includegraphics[width=11cm] {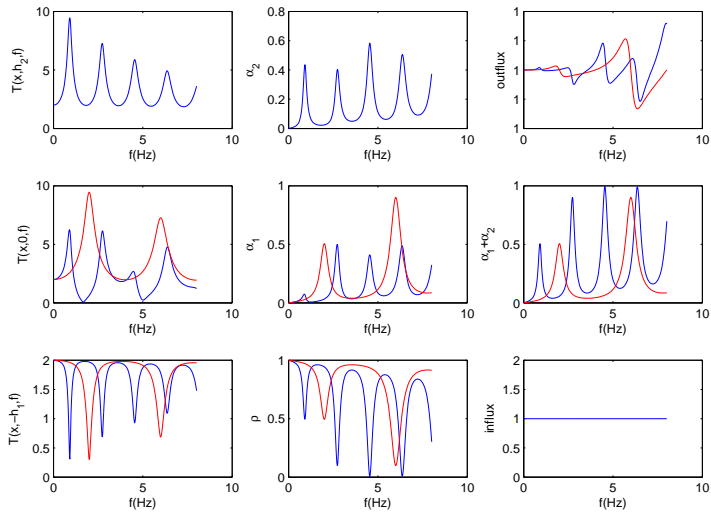}
 \caption{Same meaning of the panels as in fig. \ref{triv2}. $b^{[2]'}=200~ms^{-1}$, $b^{[2]''}=-4~ms^{-1}$, $m^{[2]}=7.2\times 10^{7}~Pa$, $h_{2}=30~m$.   Note (graphically) that the $l=1$ maximum of top response is at $0.9143~Hz$ and it amplitude is $9.449~a.u.$, the $l=3$ maximum of top response is at $2.72~Hz$ and its amplitude is $7.267~a.u.$}
  \label{triv3}
  \end{center}
\end{figure}

The case $1-g^{[21]}=0$ can correspond to two situations. The first situation is the trivial one in which the media in the overlayer and underlayer are identical,, i.e., $b^{[2]}=b^{[1]}$ and  $m^{[2]}=m^{[1]}$, in which case the configuration $\mathcal{C}_{2}$ just reduces to the configuration $\mathcal{C}_{1}$ in which the (single) layer has increased thickness $h_{3}=h_{1}+h_{2}$. Then the maxima of response are shifted to lower frequencies because $f_{2l}<f_{1l}$ without there being a change in their amplitudes at the top of the layer since
\begin{equation}\label{5-090}
\mathcal{T}_{2}(x,h_{2},\omega_{2l})=\Big\|\frac{u_{2}^{[2]}(x,h_{2};\omega_{2l})}{A^{[0]+}(\omega_{2l})}\Big\|\approx2\Big\|D_{2}^{-1}(\omega_{2l})\Big\|=
\Big\|\frac{2}{g^{[10]}}\Big\|~,
\end{equation}
wherein $\omega_{2l}=2\pi f_{2l}$. Furthermore, as concerns ground motion:
\begin{multline}\label{5-100}
\mathcal{T}_{2}(x,0,\omega_{2l})=\Big\|\frac{u_{2}^{[2]}(x,0;\omega_{2l})}{A^{[0]+}(\omega_{2l})}\Big\|\approx 2\Big\|D_{2}^{-1}(\omega_{2l})\Big\|
\Big\|\cos(k_{z}^{[2]}h_{2})\Big|_{\omega=\omega_{2l}}\Big\|\approx \\
\Big\|\frac{2}{g^{[10]}}\Big\| \Big\|\cos(k_{z}^{[2]}h_{2}+k_{z}^{[1]}h_{1}-k_{z}^{[1]}h_{1})\Big|_{\omega=\omega_{2l}}\Big\|\approx
\Big\|\frac{2}{g^{[10]}}\Big\|\Big\|\sin(k_{z}^{[1]}h_{1})\Big|_{\omega=\omega_{2l}}\Big\|
~,
\end{multline}
which, combined with (\ref{2-085}),
\begin{equation}\label{5-110}
\mathcal{T}_{1}(x,h_{2},\omega_{1l})=\Big\|\frac{u_{1}^{[1]}(x,0;\omega_{1l})}{A^{[0]+}(\omega_{1l})}\Big\|\approx\Big\|\frac{2}{g^{[10]}}\Big\|~,
\end{equation}
yields the inequality
\begin{equation}\label{5-120}
\mathcal{T}_{2}(x,h_{2},\omega_{2l})\le \mathcal{T}_{1}(x,h_{2},\omega_{1l})~;~l=1,3,5,....~,
\end{equation}
or, in other words, {\it the maximum amplitudes of ground response in the two-layer configuration in which the two layers have identical constitutive properties, are generally inferior to those in the one-layer configuration}.  These two predictions are confirmed in fig. \ref{triv2} (for $h_{2}=15~m$) and fig. \ref{triv3}  (for $h_{2}=30~m$). Note that these figures also refer to the evolution of entities (spectral absorptances and reflectances) whose definition and use are postponed for a later section.
%
%
A comment is here in order concerning what is observed in the (2,1) panel of fig. \ref{triv2}, namely the positions of the blue and red peaks near $6.5~Hz$. More precisely,  the blue peak (for the two layer configuration) is situated at higher frequency than the red peak (for the one-layer configuration) which fact seems to contradict what was stated previously. This paradoxical frequency shift is actually not abnormal because the blue peak corresponds to $f_{23}$ whereas the red peak corresponds to $f_{12}$ and since (\ref{5-080}) applies only to the same value of $l$ there is no contradiction as is apparent by the fact that the $f_{22}$ peak near $4~Hz$ is effectively located at lower frequency than the $f_{12}$ peak near $6.5~Hz$.

A second comment  concerns the (2,1) panel of fig. \ref{triv3} in which is observed   what some authors \cite{br04,ks06} term 'splitting' of the 'resonance peaks' of the one-layer configuration due to the presence of the overlayer (which in \cite{br04,ks06,sb16} is a periodic  structure idealization of a city rather than a layer). For instance, in \cite{ks06} is written "... the large fundamental peak tends to split into multiple lower peaks because of the multiple interactions between buildings. The  coincidence of the resonance frequencies between the buildings
and the soil favor these interactions", and in \cite{br04} appears the phrase "In the frequency domain, the most notable perturbations occur when the building and layer resonance frequencies coincide..., or are close. In that case, instead of a simple amplification peak, a double peak of lesser amplitude appears in the transfer function." We, on the other hand, show that it is not necessary to invoke hypotheses about the existence of multiple interactions between buildings nor phenomena due to lifting of modal resonance frequency degeneracy since neither the interaction between buildings nor the notion of resonant modes (be them of the buildings or of the site) are present in our model ($\mathcal{C}_{2}$) of the homogenized layer above the $\mathcal{C}_{1}$ site.

The second non-trivial situation in which $1-g^{[21]}=0$ requires,  for near-normal incidence and under assumption (\ref{0-001}), that
\begin{equation}\label{5-150}
\frac{m^{[2]}}{m^{[1]}}\approx \frac{b^{'[2]}}{b^{'[1]}}~,
\end{equation}
but since this condition is rarely met we shall not pursue the corresponding discussion.
\subsection{Top and bottom transfer functions in the general case: effect of variation of $h^{[2]}$}
\begin{figure}[ht]
\begin{center}
\includegraphics[width=11cm] {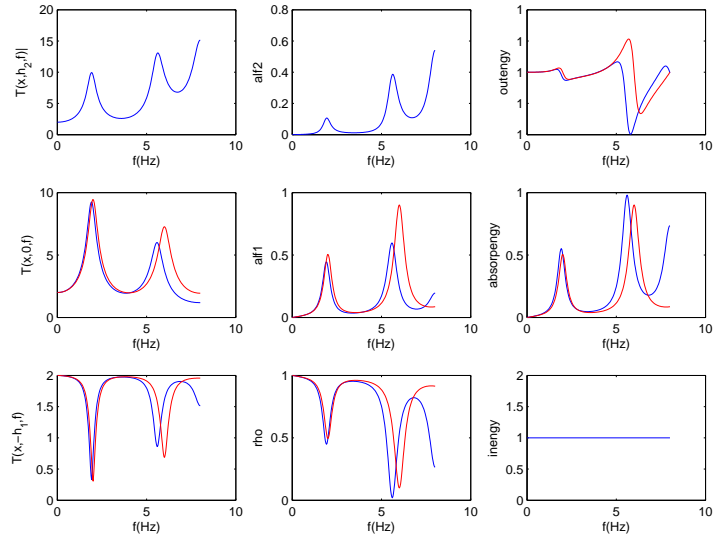}
 \caption{Same meaning of the panels as in fig. \ref{triv2}. $b^{[0]}=1000 ~ms^{-1}$, $m^{[0]}=2\times 10^{9}~Pa$, $b^{[1]}=200 ~ms^{-1}$, $b^{[1]''}=-4~ms^{-1}$, $m^{[1]}=7.2\times 10^{7}~Pa$, $b^{[2]'}=240~ms^{-1}$, $b^{[2]''}=-12~ms^{-1}$, $m^{[2]}=1.44\times 10^{7}~Pa$, $h_{1}=25~m$, $h_{2}=7.5~m$, $\theta^{i}=0^{\circ}$.   }
  \label{nonzero1}
  \end{center}
\end{figure}

\begin{figure}[ptb]
\begin{center}
\includegraphics[width=11cm] {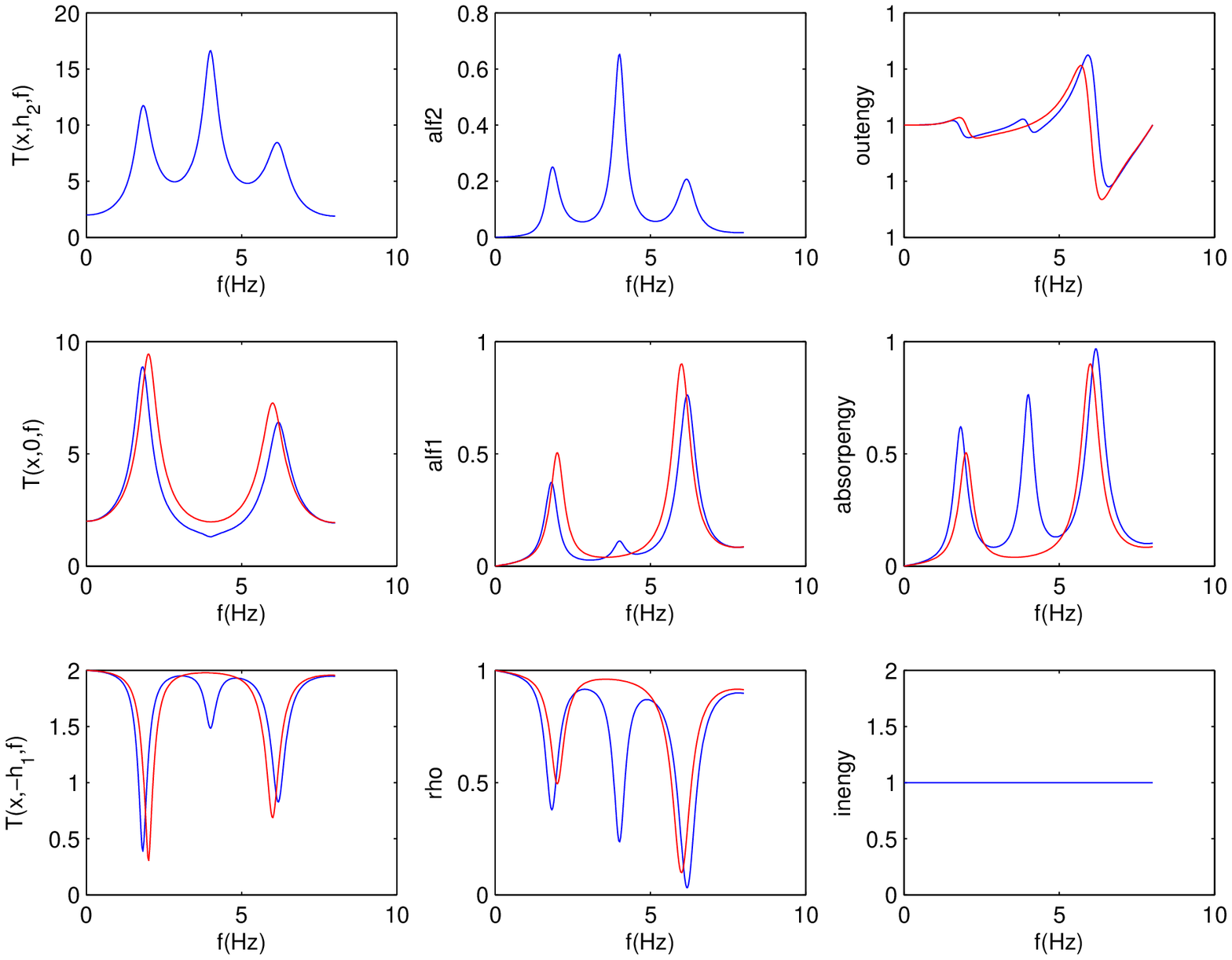}
 \caption{Same meaning of the panels as in fig. \ref{triv2}. Same parameters as in fig.  \ref{nonzero1} except that now $h_{2}=15~m$.}
  \label{nonzero2}
  \end{center}
\end{figure}

\begin{figure}[ptb]
\begin{center}
\includegraphics[width=11cm] {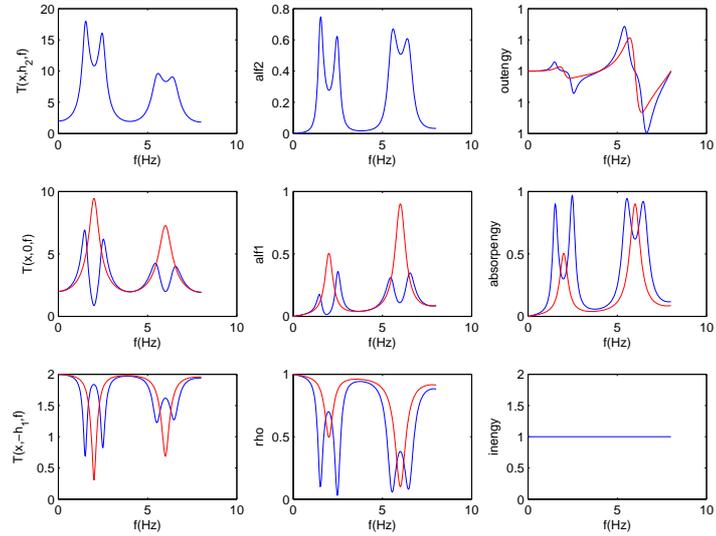}
 \caption{Same meaning of the panels as in fig. \ref{triv2}. Same parameters as in fig.  \ref{nonzero1} except that now $h_{2}=30~m$.}
  \label{nonzero3}
  \end{center}
\end{figure}

In the general case in which $1-g^{[21]}\ne0$ we are only able to obtain numerical results. Examples of the latter, for increasing $h_{2}$, are plotted in figs. \ref{nonzero1}-\ref{nonzero3}. Note the appearance of "splittings" \cite{sb16}, especially for the larger $h^{[2]}$, which were previously obtained also in the case $1-g^{[21]}= 0$.
 These figures show that it is difficult to obtain a clear picture of the effect of varying a key parameter (here $h_{2}$) of the city on entities such as (the spectral properties of) the top and bottom transfer functions. For this reason,  we now shift our attention from the spectral properties of transfer functions at two locations (top and bottom of the overlayer) to other entities that we believe are more informative about SSI because they englobe the response throughout the city (i.e., not only at two or three points).
\subsection{Can we have faith in the predictions of the layer model of  seismic response in a city?}
\begin{figure}[ht]
\begin{center}
\includegraphics[width=12cm] {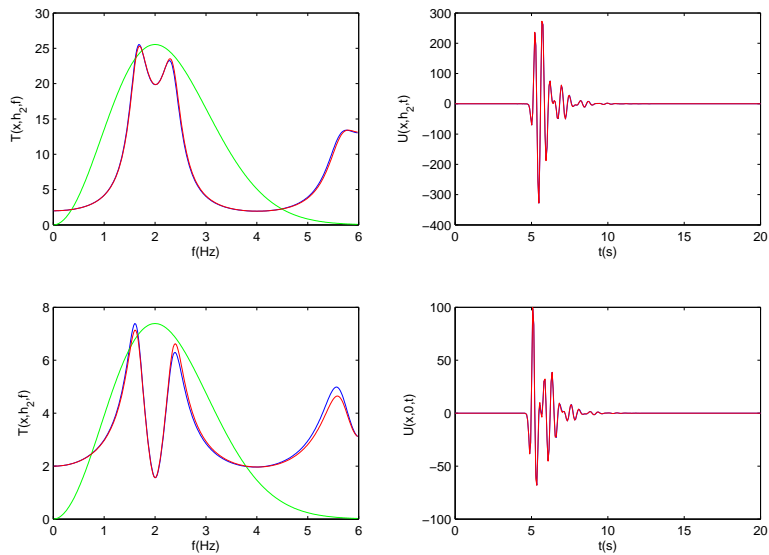}
 \caption{The upper left-hand panel depicts the top  transfer function and the upper right-hand panel  the corresponding top response signal, the lower left panel depicts the bottom (i.e., ground level, block base) transfer function and the lower  right-hand panel  the  corresponding bottom response signal. Case $\tau=5~s$, $h_{2}=30~m$, $w=10~m$, $d=20~m$ corresponding to $\phi=0.50$ and $m^{[2]}=0.50\mu^{[2]}$.}
  \label{uf-2}
  \end{center}
\end{figure}
The preceding figures concerning seismic response in cities were based on the hypothesis that the layer model could predict this response faithfully for actual cities (more precisely, for periodic block cities). That this is true, at least for top and bottom transfer functions and signals, can be seen in, for example, fig. \ref{uf-2}. Other computations not shown here indicate that the layer model predictions of these functions and signals are quite close to their rigorous counterparts relative to the periodic block model \cite{g05,gw08} as long as the city density $\phi$ is $\ge 0.5$ and the frequency does not exceed approximately $6~Hz$ for cities of the type considered in \cite{ks06} on which we shall henceforth concentrate our attention.
\section{An entity that is conserved}
 Demonstrations, such as those  given in \cite{gtw05,ks06,gw08}, of what becomes of seismic motion on the bottom (ground level) and  top of the city or its surrogate (i.e., the overlayer) are not necessarily transposable to the overall motion  and energy within the overlayer. We know empirically that some of the buildings  will be damaged or destroyed during earthquakes that hit a city which means that they are necessarily the recipients of a part of the incident energy. How, and how much of, this incident energy is injected into the buildings is the as-yet unanswered question.

 After having established expressions for the fields in the various subdomains of the configuration $C_{II}$ or $C_{2}$, the next step towards an answer to this question is to establish an expression for the conservation of a certain entity tied up with the energy and which we call 'flux', relating the incident flux to the fluxes  distributed to the component areas of the scattering configuration.
\begin{figure}[ht]
\begin{center}
\includegraphics[width=0.65\textwidth]{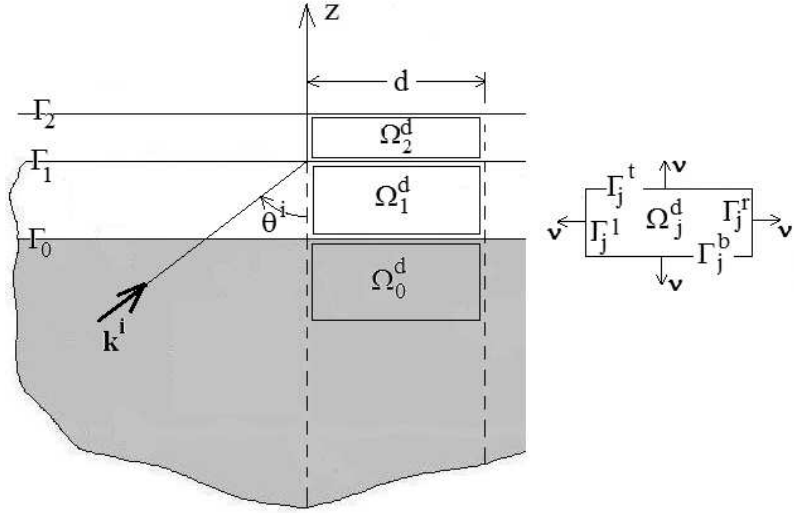}
\caption{Sagittal plane view of  the integration  domains $\Omega_{j}^{d}~;~1,2,3$ for establishment of the conservation of flux (and later energy) relation.}
\label{consvengy}
\end{center}
\end{figure}
In fig. \ref{consvengy}, we depict the rectangular integration domains $\Omega_{j}^{d}~;~0,1,2$ whose boundaries are $\partial\Omega_{j}^{d}=\Gamma_{j}^{l}\cup\Gamma_{j}^{b}\cup\Gamma_{j}^{r}\cup\Gamma_{j}^{t}~;~0,1,2$. Each of the horizontal segments are of length $d$, and the lengths of $\Gamma_{j}^{l}~;~0,1,2$ and $\Gamma_{j}^{r}~;~0,1,2$ are $h_{j}$, it being understood that we shall take the limit $h_{0}\rightarrow\infty$. Note that  $\Omega_{2}^{d}$ is the domain of a generic block if the city is considered to be composed of a periodic set of identical blocks,
\begin{equation}\label{4-010}
\Gamma_{0}^{t}=\Gamma_{1}^{b}\subset\Gamma_{0}~~,~~\Gamma_{1}^{t}=\Gamma_{2}^{b}\subset \Gamma_{1}~~,~~\Gamma_{2}^{t}\subset \Gamma_{2}~,
\end{equation}
and that the outward-pointing unit normal from $\Omega_{j}^{d}~;~0,1,2$  is $\boldsymbol{\nu}$. We are treating configuration $\mathcal{C}_{2}$ (although the generalization to the other configurations is obvious) and adopt the simplified notations $u^{[j]}=u_{2}^{[j]}=u^{[j]}$ and $u^{[j]\pm}=u_{2}^{[j]\pm}$.

Let us first recall the governing equations of our  layer-like city problem (whose solution was expressed in (\ref{1-010}), (\ref{1-015}), (\ref{1-110}), (\ref{1-125}) and (\ref{1-130})):
\begin{equation}\label{1-020}
u_{,xx}^{[l]}(\mathbf{x},\omega)+u_{,zz}^{[l]}(\mathbf{x},\omega)+(k^{[l]})^{2}u^{[l]}(\mathbf{x},\omega)=0~;~\mathbf{x}\in \Omega_{l}~;~l=0,1,2~,
\end{equation}
\begin{equation}\label{1-030}
m^{[1]}u_{,z}^{[1]}(\mathbf{x},\omega)=0~;~\mathbf{x}\in \Gamma_{2}~,
\end{equation}
\begin{equation}\label{1-040}
u^{[l]}(\mathbf{x},\omega)-u^{[1+1]}(\mathbf{x},\omega)=0~;~\mathbf{x}\in \Gamma_{l}~;~l=0,1~,
\end{equation}
\begin{equation}\label{1-050}
m^{[l]}u_{,z}^{[l]}(\mathbf{x},\omega)-m^{[1+]}u_{,z}^{[1+1]}(\mathbf{x},\omega)=0~;~\mathbf{x}\in \Gamma_{l}~;~l=0,1~,
\end{equation}
\begin{equation}\label{1-060}
u^{[0]-}(\mathbf{x},\omega)\sim \text{outgoing waves}~;~\mathbf{x}\rightarrow\infty~,
\end{equation}
wherein   $u_{,\zeta}$ ($u_{,\zeta\zeta}$) denotes the first (second) partial derivative of $u$ with respect to $\zeta$.

 Eq. (\ref{1-020}) leads to
\begin{equation}\label{4-020}
\int_{\Omega_{j}^{d}}[u^{[j]*}\Delta u^{[j]}-u^{[j]}\Delta u^{[j]*}]d\varpi+2i\Im\big[(k^{[j]})^{2}\big]\int_{\Omega_{j}^{d}}\|u^{[j]}\|^{2}d\varpi=0~;~0,1,2~,
\end{equation}
wherein the symbol $^{*}$ designates the complex conjugate, and $d\varpi$  the differential area element. Green's second identity tells us that
\begin{equation}\label{4-030}
\int_{\Omega_{j}^{d}}[u^{[j]*}\Delta u^{[j]}-u^{[j]}\Delta u^{[j]*}]d\varpi=\int_{\partial\Omega_{j}^{d}}[u^{[j]*}\boldsymbol{\nu}\cdot\nabla u^{[j]}-u^{[j]}\boldsymbol{\nu}\cdot\nabla u^{[j]*}]d\gamma~;~0,1,2
~,
\end{equation}
wherein $d\gamma$ is the differential boundary element and $\boldsymbol\nu$ the unit vector pointing outwards from the boundary of $\Omega_{j}^{d}$, so that
\begin{equation}\label{4-040}
I^{[j]}=\Im\int_{\partial\Omega_{j}^{d}}u^{[j]*}\boldsymbol{\nu}\cdot\nabla u^{[j]}d\gamma=-\Im\big[(k^{[j]})^{2}\big]\int_{\Omega_{j}^{d}}\|u^{[j]}\|^{2}d\varpi=-J^{[j]}~;~0,1,2
~.
\end{equation}
We note that $J^{[0]}=0$ because it was  assumed that $\Im \beta^{[0]}=0$. Moreover,
\begin{multline}\label{4-050}
I^{[j]}=\Im\int_{\Gamma_{j}^{l}}u^{[j]*}\boldsymbol{\nu}\cdot\nabla u^{[j]}d\gamma+\Im\int_{\Gamma_{j}^{b}}u^{[j]*}\boldsymbol{\nu}\cdot\nabla u^{[j]}d\gamma+\\
\Im\int_{\Gamma_{j}^{r}}u^{[j]*}\boldsymbol{\nu}\cdot\nabla u^{[j]}d\gamma+\Im\int_{\Gamma_{j}^{t}}u^{[j]*}\boldsymbol{\nu}\cdot\nabla u^{[j]}d\gamma=
 I^{[j]l}+I^{[j]b}+I^{[j]r}+I^{[j]t}~;~0,1,2
~.
\end{multline}
Eqs. (\ref{1-015})  implies
\begin{equation}\label{4-060}
u^{[j]}(x+d,z;\omega)=u^{[j]}(x,z;\omega)\exp(ik_{x}d)~,
\end{equation}
whatever is the chosen value of $d$, so that
\begin{equation}\label{4-070}
I^{[j]l}+I^{[j]r}=0~;~j=0,1,2
~.
\end{equation}
Similarly, by applying the boundary conditions (\ref{1-030})-(\ref{1-050}) and making use of the fact that the $m^{[j]}$ were all assumed to be real,  we find:
\begin{equation}\label{4-080}
I^{[0]t}=-\frac{m^{[1]}}{m^{[0]}}I^{[1]b}~~,~~I^{[2]b}=-\frac{m^{[1]}}{m^{[2]}}I^{[1]t}~~,~~I^{[2]t}=0
~,
\end{equation}
whence
\begin{equation}\label{4-090}
I^{[0]}=I^{[0]b}-\frac{m^{[1]}}{m^{[0]}}I^{[1]b}~~,~~I^{[1]}=I^{[1]b}+I^{[1]t}~~,~~I^{[2]}=-\frac{m^{[1]}}{m^{[2]}}I^{[1]t}
~.
\end{equation}
But $J^{[0]}=0$ means that $I^{[0]}=0$ so that $(\ref{4-090})$ and (\ref{4-040}) entail
\begin{equation}\label{4-110}
\frac{m^{[1]}}{m^{[0]}}J^{[1]}+\frac{m^{[2]}}{m^{[0]}}J^{[2]}=-I^{[0]b}
~.
\end{equation}
this being the basic form of the conservation of flux relation \cite{p95} in which it is implicit that $h_{0}\rightarrow \infty$.

After further manipulations we find
\begin{equation}\label{4-190}
\frac{m^{[1]}}{m^{[0]}}J^{[1]}+\frac{m^{[2]}}{m^{[0]}}J^{[2]}=k_{z}^{[0]}d\|A^{[0]+}\|^{2}-k_{z}^{[0]}d\|A^{[0]-}\|^{2}
~,
\end{equation}
or, generalized to the three configurations $\mathcal{C}_{m}~;~m=0,1,2$:
\begin{equation}\label{4-200}
\rho_{m}(\omega)+\sum_{j=1}^{m}\alpha_{m}^{[j]}(\omega)=1~;~m=1,2,~~,~~\rho_{0}(\omega)=1~,
\end{equation}
wherein
\begin{equation}\label{4-210}
\rho_{m}(\omega)=\frac{\|A_{m}^{[0]-}\|^{2}}{\|A^{[0]+}\|^{2}}~~,~~\alpha_{m}^{[j]}(\omega)=
\frac{1}{k_{z}^{[0]}d}\frac{m^{[j]}}{m^{[0]}}\Im\big[(k^{[j]})^{2}\big]\int_{\Omega_{j}^{d}}\Big\|\frac{u_{m}^{[j]}}{A^{[0]+}}\Big\|^{2}d\varpi,
\end{equation}
 with $\rho_{m}(\omega)$  the {\it spectral reflectance} \cite{p95} (in the half space, this being associated with what is often termed  "radiation damping"  \cite{ws96}), and  $\alpha_{m}^{[j]}$ the {\it spectral absorptance} \cite{p95} in the domain $\Omega_{j}^{d}$ occupied by medium $\mathcal{M}^{[j]}$, it being recalled that $\mathcal{M}^{[0]}$ is non-dissipative. Eq. (\ref{4-200}) (for $m=2$) signifies that {\it the larger is $\alpha_{2}^{[2]}$, the more of the input flux is transferred into the overlayer (and the less is the fraction of input flux transferred to the underlayer and/or to radiation damping), this being a consequence of the the conservation of flux}.

 More specifically, (\ref{4-200}), whose  right-hand term is the normalized input flux (furnished by the incident seismic wave) and left-hand term the normalized output flux, is the rigorous, developed, expression of the conservation of flux for the layer model $\mathcal{C}_{2}$ of the city (for $m=2$, and even for the two other configurations). Eqs. (\ref{4-200})-(\ref{4-210}) are rigorously applicable as such also to the situation in which the city is in the form of a periodic set of identical blocks, in which case $\Omega_{2}^{d}$ is the domain occupied by a generic block.

A last word on these conservation expressions in the case of the layer model city: the appearance of $d$ therein seems to indicate that they depend on the width of the spatial integration domains. That this is  not true can be seen as follows. The displacement in the $j$-th domain is of the form $u_{2}^{[j]}(\mathbf{x};\omega)=w_{2}(z;\omega)\exp(ik_{x}x)$ so that (with $z_{1}^{-}=-h_{1}$, $z_{1}^{+}=z_{2}^{-}=0$, $z_{2}^{+}=h_{2}$)
\begin{equation}\label{4-260}
\frac{1}{d}\int_{\Omega_{j}^{d}}\|u_{2}^{[j]}\|^{2}d\varpi=
\frac{1}{d}\int_{z_{-}}^{z_{+}}dz\|u_{2}^{[j]}\|^{2}\int_{0}^{d}dx=
\int_{z_{j}^{-}}^{z_{j}^{+}}dz\|u_{2}^{[j]}\|^{2}~,
\end{equation}
which does not depend on $d$. In the case of the periodic block model of the city, the conservation of flux relation does, however, depend on the period $d$.
\section{Theoretical and numerical predictions of how the variations of the city height, city density and/or block shear modulus affect the behavior of the overlayer spectral absorptance and the two transfer fucntions}\label{matt}
%
\subsection{Theoretical predictions deriving from the layer model}\label{tmatt}
We are here concerned exclusively with configuration $\mathcal{C}_{2}$, and, in particular, with the two transfer functions $\mathcal{T}_{2}(x,h_{2},\omega)$,
$\mathcal{T}_{2}(x,0,\omega)$ and the overlayer spectral absorptance $\alpha_{2}(\omega)$.

At first, we drop the assumption (\ref{0-001}) so that $k^{[j]}~;~j=1,2$ are complex which also implies that $k_{z}^{[j]}~;~j=1,2$ are complex, i.e.,  $k_{z}^{[j]}=k_{z}^{'[j]}+i k_{z}^{''[j]}$. By means of the expressions for $u_{2}^{[2]}$ and $A^{[2]\pm}$ we find:
\begin{equation}\label{7-010}
\mathcal{T}_{2}(x,h_{2},\omega)=\frac{2}{\|D_{2}\|}~,
\end{equation}
\begin{equation}\label{7-020}
\mathcal{T}_{2}(x,0,\omega)=
\frac{2
\left|\sqrt{
\left(
\cos(k_{z}^{'[2]}h_{2})
\cosh(k_{z}^{''[2]}h_{2})
\right)^{2}
+\left(\sin(k_{z}^{'[2]}h_{2})\sinh(k_{z}^{''[2]}h_{2})\right)^{2}}\right|
}
{\|D_{2}\|}~,
\end{equation}
We also find, by explicit integration and some algebraic manipulations:
\begin{equation}\label{7-030}
\alpha_{2}(\omega)=\frac{1}{\|D_{2}\|^{2}}\frac{g^{[20]}}{k_{z}^{[2]}}\left[k_{z}^{'[2]}\sinh(2k_{z}^{''[2]}h_{2})+k_{z}^{''[2]}\sin (2k_{z}^{'[2]}h_{2})\right]
~,
\end{equation}
The second step consists in the re-invocation of (\ref{0-001}) which, knowing that sinc$(\zeta)=\frac{\sin\zeta}{\zeta}$ and $\eta^{[j]}=\sqrt{\big(\frac{b^{[0]}}{b^{'[j]}}\big)^{2}-\big(s^{i}\big)^{2}}$, entails
\begin{equation}\label{7-040}
g^{[jk]}\approx \tilde{g}^{[jk]}=\frac{m^{[j]}\eta^{[k]}}{m^{[k]}\eta^{[j]}}~~,~~\alpha_{2}(\omega)\approx \tilde{\alpha}_{2}(\omega)=\frac{\tilde{g}^{[20]}}{\|\tilde{D}_{2}\|^{2}}
2k^{0}h_{2}k_{z}^{''[2]}\left[1+\text{sinc}\left(2k_{z}^{'[2]}h_{2}\right)\right]
~,
\end{equation}
\begin{equation}\label{7-050}
\mathcal{T}_{2}(x,h_{2},\omega)\approx\mathcal{\tilde{T}}_{2}(x,h_{2},\omega)=\frac{2}{\|\tilde{D}_{2}\|}~~,~~
\mathcal{T}_{2}(x,0,\omega)\approx\mathcal{\tilde{T}}_{2}(x,0,\omega)=
\frac{2\left\|\tilde{C}^{[22]}\right\|}{\|\tilde{D}_{2}\|}~,
\end{equation}
\begin{equation}\label{7-070}
D_{2}\approx\tilde{D}_{2}=\left(\tilde{C}^{[22]}\tilde{C}^{[11]}-\tilde{g}^{[21]}\tilde{S}^{[22]}\tilde{S}^{[11]}\right)-
i\tilde{g}^{[10]}\left(\tilde{C}^{[22]}\tilde{S}^{[11]}+\tilde{g}^{[21]}\tilde{S}^{[22]}\tilde{C}^{[11]}\right)
~,
\end{equation}
\begin{equation}\label{7-080}
C^{[jj]}\approx\tilde{C}^{[jj]}=\cos\left(k_{z}^{'[j]}h_{j}\right)~,~S^{[jj]}\approx\tilde{S}^{[jj]}=\sin\left(k_{z}^{'[j]}h_{j}\right)
~.
\end{equation}
In all our preceding examples, whose parameters are based mostly on those in \cite{ks06}, the incident angle was $\theta^{i}=0^{\circ}$, the underlayer thickness  $h_{1}$ was fixed and taken to be $25~m$, the shear wavespeed was also fixed and taken to be $b_{1}=1000~ms^{-1}$ and the Ricker pulse characteristic frequency $\nu$ was usually taken to be $2~Hz$. In addition, we assumed, and continue to assume, that:  $b^{[0]'}=1000 ~ms^{-1}$, $b^{[0]''}= 0~ms^{-1}$, $m^{[0]}=2\times 10^{9} Pa$, $b^{[1]'}=200 ~ms^{-1}$, $b^{[1]''}=-4~ms^{-1}$, $m^{[1]}=7.2\times 10^{7} Pa$, $b^{[2]'}=240~ms^{-1}$, $b^{[2]''}=-12~ms^{-1}$, $\mu^{[2]}=1.44\times 10^{7} Pa$.

With these fixed choices of $h_{1}$, $b^{[1]}$ and frequency  (the last one for $f=\nu=2~Hz$), $C^{[11}$ and $S^{[11}$ will be fixed, which incites us to re-cast $D_{2}$ in the form
\begin{multline}\label{7-090}
D_{2}=C^{[11]}\left(C^{[22]}-ig^{[20]}S^{[22]}\right)-iS^{[11]}\left(g^{[10]}C^{[22]}+ig^{[21]}S^{[22]}\right)\approx\\
\tilde{D}_{2}=\tilde{C}^{[11]}\left(\tilde{C}^{[22]}-i\tilde{g}^{[20]}\tilde{S}^{[22]}\right)-
i\tilde{S}^{[11]}\left(\tilde{g}^{[10]}\tilde{C}^{[22]}+i\tilde{g}^{[21]}\tilde{S}^{[22]}\right)
~.
\end{multline}
The previously-cited parameter values entail $k_{z}^{'[1]}h_{1}=\pi/2$ so that $C^{[11]}=0$, $S^{[11]}=1$ and
\begin{equation}\label{7-100}
\left\|\tilde{D}_{2}\right\|^{-1}=\left\|\tilde{g}^{[10]}\tilde{C}^{[22]}+i\tilde{g}^{[21]}\tilde{S}^{[22]}\right\|^{-1}
.
\end{equation}
Adopting, as usual, the value $240~ms^{-1}$ for the shear wavespeed $b^{[2]}$ in the (buildings of the) overlayer, we obtain $\tilde{C}^{[22]}=\cos\left(\pi h_{2}/60 \right)$ and $\tilde{S}^{[22]}=\sin\left(\pi h_{2}/60 \right)$, so that (recalling that the quantities with tildes are real):\\

 case (a) in which $h_{2}=0,~60~120~m,...$: $\left\|\tilde{D}_{2}\right\|^{-1}=\tilde{g}^{[01]}$    (because $\tilde{S}^{[22]}=0$) and\\

 case (b) in which $h_{2}=30,~90,~150~m,...$: $\left\|\tilde{D}_{2}\right\|^{-1}=\tilde{g}^{[12]}$  (because $\tilde{C}^{[22]}=0)$.\\
 Consequently, in case (a)
\begin{equation}\label{7-110}
\tilde{\mathcal{T}}_{2}(x,h_{2},\omega)=\tilde{\mathcal{T}}_{2}(x,0,\omega)=2\tilde{g}^{[01]}~~,~~
\tilde{\alpha}_{2}(\omega)=2\tilde{g}^{[20]}\left(\tilde{g}^{[01]}\right)^2
k^{0}h_{2}k_{z}^{''[2]}\left[1+\text{sinc}\left(2k_{z}^{'[2]}h_{2}\right)\right]
,
\end{equation}
and in case (b)
\begin{equation}\label{7-120}
\tilde{\mathcal{T}}_{2}(x,h_{2},\omega)=2\tilde{g}^{[12]}~~,~~\tilde{\mathcal{T}}_{2}(x,0,\omega)=0~~,~~
\tilde{\alpha}_{2}(\omega)=2\tilde{g}^{[10]}\tilde{g}^{[12]}
k^{0}h_{2}k_{z}^{''[2]}\left[1+\text{sinc}\left(2k_{z}^{'[2]}h_{2}\right)\right]
.
\end{equation}
From this, and the fact that only $\tilde{g}^{[jk]}$ with either $j=2$ or $k=2$  depends on  $m^{[2]}$ (and therefore on $\phi$), we can make the following predictions:\\

(a1) in case (a), the two transfer functions do not vary either with $m^{[2]}$ (and therefore with $\phi$),\\

(a2) in case (a), the overlayer spectral absorptance increses linearly with increasing $m^{[2]}$ (and therefore with increasing $\phi$) and increasing $h_{2}$,\\

(b1) in case (b), the top transfer function decreases with  increasing $m^{[2]}$ (and therefore with increasing $\phi$), whereas the bottom transfer function is nil,\\

(b2) in case (b), the overlayer absorptance decreases with increasing $m^{[2]}$ (and therefore with increasing $\phi$) and increases with increasing $h_{2}$,\\

(c1) the two transfer functions are oscillating functions of $h_{2}$ for fixed frequency, but the values of the maxima and minima of these transfer functions do not depend on $h_{2}$,\\

(c2) the overlayer spectral absorptance is an oscillating function of $h_{2}$ and the values of the maxima an minima of this function increase with increasing $h_{2}$.
\subsection{Numerical predictions for the case treated in Kham et al.}\label{nbmatt}
\begin{figure}[ht]
\begin{center}
\includegraphics[width=11.5cm] {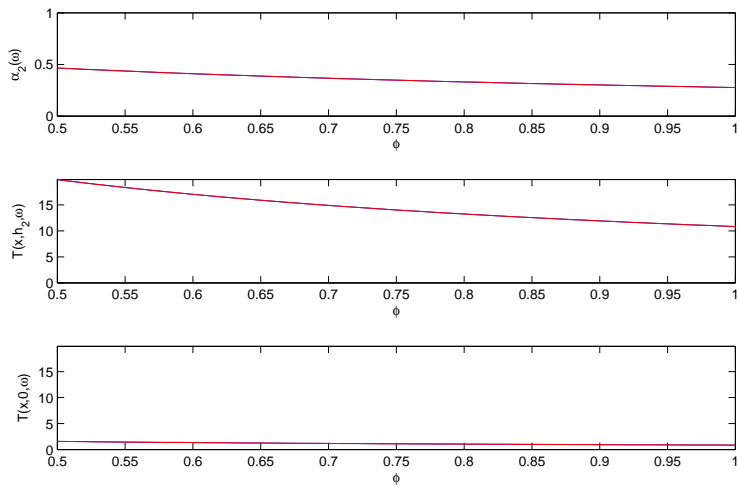}
 \caption{Illustration of predictions (b1) and (b2) with regard to increasing city density $\phi$. Top panel: $\alpha_{2}$ as a function of $\phi$. Middle panel:  $\mathcal{T}_{2}(x,h_{2},\omega)$ as a function of $\phi$. Bottom panel: $\mathcal{T}_{2}(x,0,\omega)$ as a function of $\phi$. In all panels, the blue curves apply to the periodic narrow-block model ($w=10~m$) of the city and the red curves to the corresponding overlayer model of the city. Case (b), i.e., $h_{2}=30~m$, $f=2~Hz$. }
  \label{att-01}
  \end{center}
\end{figure}

The periodic-block city to which the study of Kham et al. \cite{ks06} is devoted is of the narrow-block (actually buildings of width $w=10~m$) variety. Probably because it is uncommon to encounter buildings with aspect ratios $h/d$ that exceed 3, Kham et al. chose $h=30~m$ (and, in fact no other, significantly-larger, building heights for their investigation of how increasing city density affects seismic response in a city. Moreover, Kham et al. did not consider an entity such as the spectral absorptance, subjected to a conservation principle,   with regard to the question of whether the seismic response in the built component of denser cities is more or less pronounced than in that of less-dense cities. Thus, in order to reproduce (and possibly extend by means of $\alpha_{2}$) the results of Kham et al. and compare them to those of our layer model, we had no other option than to choose a narrow-block ($w=10~m$) city of height $h_{2}=30~m$, which falls into case (b) of sect. \ref{tmatt}. The corresponding numerical results are shown in fig. \ref{att-01}.

We observe  in this figure that the predictions (b1) of the layer model are in good agreement (the small differences are due to the fact that the theoretical predictions apply to entities with tildes $\tilde{\mathcal{T}}_{2}$, $\tilde{\alpha}_{2}$ and the computed results to the corresponding entities $\mathcal{T}_{2}$, $\alpha_{2}$ without tildes) both with our rigorous  periodic block layer model \cite{g05,gw08} and also with the findings of Kham et al. \cite{ks06} according to which increasing city density provokes a substantial decrease of the top transfer function and a small  decrease  of the bottom (i.e., ground level within the blocks) transfer function (which is practically nil). This figure also shows, in agreement with prediction (b2), that increasing city density  results in decreasing spectral absorptance in the overlayer (i.e., in what is assumed to replace the city). Thus, at this stage, everything seems to be in agreement with the conclusions in \cite{ks06} that increasing city density has a 'beneficial' effect on the seismic response therein. But,  we must pay more attention to the uppermost panel in fig. \ref{att-01} which tells us that, for the chosen frequency $f=2~Hz$, {\it from a quarter to a little less than a half of the incident seismic flux is transferred to, and spent within, the built portion of the site/city configuration} over this range of city densities, which shows that cities modify substantially, in an unfavorable manner (with respect to the situation in which they are absent and in which $\alpha_{2}$ is trivially nil) the seismic response of the  geophysical configuration in that they are the recipients of a substantial portion of an entity which, as we shall show farther on, is connected with the energy of the seismic solicitation.
\subsection{Numerical predictions for the effect of increasing den.sity of wider-block cities}\label{wbmatt}
\begin{figure}[ht]
\begin{center}
\includegraphics[width=11.5cm] {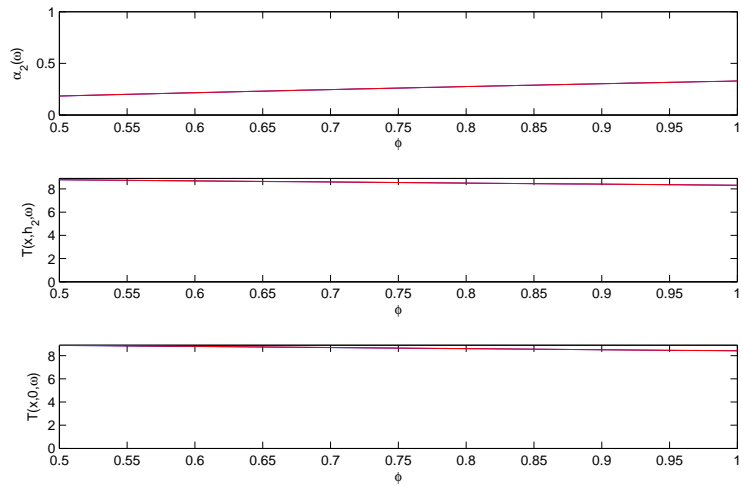}
 \caption{Illustration of predictions (a1) and (a2) with regard to increasing city density $\phi$. Same meaning of panels as in fig. \ref{att-01} except that now $w=50~m$.  Case (a), i.e., $h_{2}=60~m$, $f=2~Hz$. }
  \label{att-02}
  \end{center}
\end{figure}

By enlarging the width of the blocks we are less-prone to the aspect ratio problem evoked in the preceding section. By taking $w=50~m$, it is legitimate to consider blocks whose height is $h_{2}=60~m$ and even higher and thus find out what happens in case (a) (defined in sect. \ref{tmatt}). This is done in fig. \ref{att-02} wherein we observe, in agreement with prediction (b1), that the two transfer functions decrease with increasing city density. Moreover, in the upper panel of this figure, we see that the spectral absorptance in the overlayer {\it increases} with increasing city density, so that if we assume that $\alpha_{2}$  is a better indicator than the transfer functions of the seismic response in the built component of the city then we must conclude that increasing city density has an adverse (rather than beneficial) effect, at least for certain city heights.

\begin{figure}[ptb]
\begin{center}
\includegraphics[width=11.5cm] {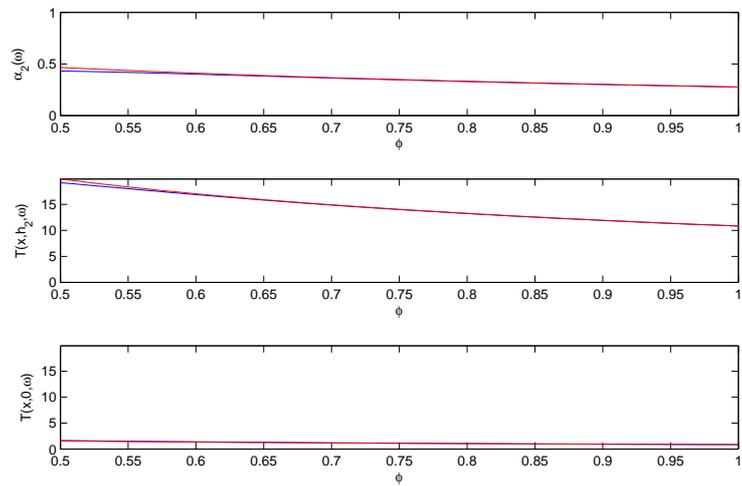}
 \caption{Illustration of predictions (b1) and (b2) with regard to increasing city density $\phi$. Same meaning of panels as in fig. \ref{att-01}  except that now $w=50~m$.  Case (b), i.e., $h_{2}=30~m$,  $f=2~Hz$.}
  \label{att-03}
  \end{center}
\end{figure}

In fact, when the city height is brought back to $h_{2}=30~m$, this wide-block city responds in the same manner as the narrow-block city of the previous section and predictions (b1) and (b2) are again verified, as seen in fig. \ref{att-03}.
\clearpage
\newpage
\subsection{Numerical predictions for the effect of increasing city height}\label{hmatt}
Now, since we are going to consider rather tall cities and  want to stay within the framework of reasonable aspect-ratio blocks, we again choose the width of the latter to be $w=50~m$.
\begin{figure}[ht]
\begin{center}
\includegraphics[width=11.5cm] {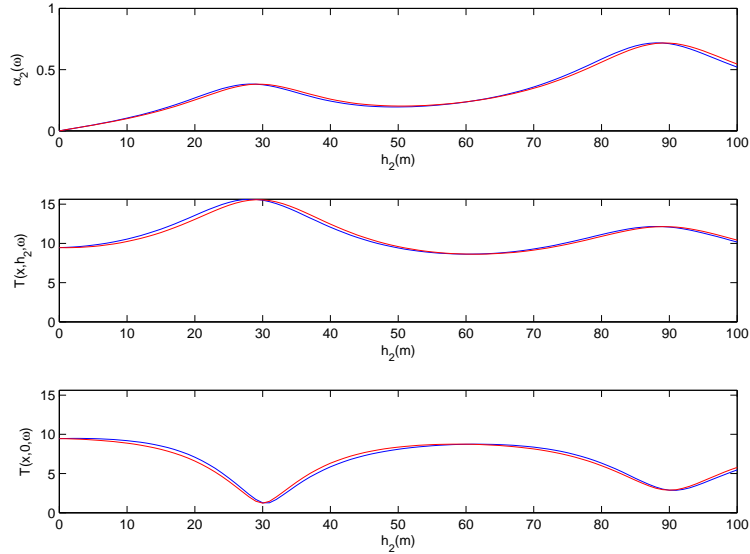}
 \caption{Illustration of predictions (c1) and (c2) with regard to increasing city height $h_{2}$.  Top panel: $\alpha_{2}$ as a function of $h_{2}$. Middle panel:  $\mathcal{T}_{2}(x,h_{2},\omega)$ as a function of $h_{2}$. Bottom panel: $\mathcal{T}_{2}(x,0,\omega)$ as a function of $h_{2}$. In all panels, the blue curves apply to the periodic block model  of the city and the red curves to the corresponding overlayer model of the city.    $w=50~m$, $d=75~m$ so that $\phi=0.67$.  $f=2~Hz$.}
  \label{hatt-01}
  \end{center}
\end{figure}
In fig. \ref{hatt-01}, we can notice the good agreement of the layer model with periodic block model numerical results. Moreover, the layer model predictions (c1) and  (c2) are well-verified. Note that whether the oscillating response  corresponds  to a maximum or minimum depends in cases (b) and (a)  on the value of $g^{[12]}$ in relation  to the value of $g^{[01]}$. Note that $\sim 70\%$ of the incident seismic flux is transferred to the built component of the city when its height is $90 ~m$, which  (again) indicates danger for the buildings of a city submitted to an earthquake.
\section{Energy considerations}
Another commonly-employed indicator of SSI in connection with the specific class of above-ground structures that are cities, is the kinetic energy at specific points of the structure. For instance, in \cite{gtw05}, quasi-rigorous computations are made of the kinetic energy integrated over the duration of the response signal at the midpoints of the top segment of the generic block and of the segment on the ground between successive buildings, normalized by these energies in the absence of the blocks.  It is shown in \cite{gtw05} that these so-called 'vulnerability indices' can attain significantly-large values ($>1$), indicative of large-amplitude, large duration shaking of both the buildings and the ground during an earthquake. In \cite{ks06}, computations are made, for various city densities, of a similar type of normalized integrated-over the signal duration-kinetic energy, whereby  Kham et al.  find, in their fig. 6, that it: (i) decreases with increasing city density and (b) is always $<1$. These findings  appear to be consistent with what they found concerning the behavior of the ground transfer function.

The question is whether such a 'ground kinetic energy' or 'ground vulnerability index' is something that  is really informative of the way energy is injected into the city. In fact, it would seem that an entity that has only to do with response on the ground does not necessarily tell us how much energy is transferred into the buildings, which energy is spent in eventually damaging or destroying the buildings, so that finding that ground motion  is less for denser cities does not necessarily mean that the damage inflicted to the buildings of denser cities is less than that inflicted to the buildings of less dense cities. Also, the discussions in \cite{gtw05,ks06} deal neither with the amount of energy  absorbed in the layer (or layer-like basin) beneath the ground, nor with radiation damping (i.e., energy spent by waves sent back into the lower half space), nor, for this reason, with the question of whethe energy is conserved.

It is important to understand that although (\ref{4-200}) expresses a conservation principle, it takes no account of the spectrum of the incident seismic pulse. Since this spectrum (just like the signal duration in the time domain) obviously conditions the amount of energy injected into the city it must play a central role in energy computations. The inclusion of the solicitation spectrum is done in the next section, in a manner by which the conservation principle is preserved.
\subsection{Frequency-integrated form of the conservation of flux leading to the conservation of energy relation}
We re-write (\ref{4-200}) (for $\mathcal{C}_{2}$) as
\begin{equation}\label{4-230}
k_{z}^{[0]}m^{[0]}\|A^{[0]-}\|^{2}+\frac{m^{[1]}}{d}\Im\big[(k^{[1]})^{2}]\int_{\Omega_{1}^{d}}\|u_{2}^{[1]}\|^{2}d\varpi+
\frac{m^{[2]}}{d}\Im\big[(k^{[2]})^{2}]\int_{\Omega_{2}^{d}}\|u_{2}^{[2]}\|^{2}d\varpi=k_{z}^{[0]}m^{[0]}\|A^{[0]+}\|^{2}.
\end{equation}
This expression is then integrated over all angular frequencies:
\begin{multline}\label{4-240}
\int_{0}^{\infty}d\omega k_{z}^{[0]}m^{[0]}\|A^{[0]-}\|^{2}+\frac{m^{[1]}}{d}\int_{0}^{\infty}d\omega\Im\big[(k^{[1]})^{2}]\int_{\Omega_{1}^{d}}d\varpi\|u_{2}^{[1]}\|^{2}+\\
\frac{m^{[2]}}{d}\int_{0}^{\infty}d\omega\Im\big[(k^{[2]})^{2}]\int_{\Omega_{2}^{d}}d\varpi\|u_{2}^{[2]}\|^{2}=
\int_{0}^{\infty}d\omega k_{z}^{[0]}m^{[0]}\|A^{[0]+}\|^{2}~.
\end{multline}
Recall that $k^{[j]}=\omega/b^{[j]}=\omega r^{[j]}/m^{[j]}$ so that by virtue of (\ref{0-001})
\begin{equation}\label{4-250}
\Im[(k^{[j]})^{2}]=
\frac{-2\omega^{2}b^{'[j]}b^{''[j]}}{\|(b^{[j]})^{2}\|^{2}}\approx ~-2\omega^{2}\frac{b^{''[j]}}{b^{'[j]}}\frac{r^{'[j]}}{m^{[j]}}~,
\end{equation}
whence
\begin{equation}\label{4-260}
\frac{m^{[j]}}{d}\int_{0}^{\infty}d\omega\Im\big[(k^{[j]})^{2}]\int_{\Omega_{j}^{d}}d\varpi\|u^{[1]}\|^{2}\approx
\frac{2}{d}\Big(\frac{-b^{''[j]}}{b^{'[j]}}\Big)\left[\int_{\Omega_{j}^{d}}d\varpi
r^{'[j]}\int_{0}^{\infty}d\omega\omega^{^{2}}\|u^{[j]}\|^{2}\right]~,
\end{equation}
The term $[~~]$ can be recognized, via Parseval's theorem, to be the total kinetic energy in the layer occupying $\Omega_{j}^{d}$. From this, we can conclude that (\ref{4-240}) genuinely expresses conservation of energy for our $\mathcal{C}_{2}$ (or $\mathcal{C}_{II}$)  configuration. Note that the first term on the left-hand side therein represents the reflected energy, the second and third terms the absorption due to dissipation of kinetic energy in the underlayer and overlayer (or generic city block) respectively  and the term on the right-hand side represents the input energy furnished by the incident seismic body wave.

The conservation of energy relation  can be written as
\begin{equation}\label{4-270}
\mathcal{R}+\sum_{j=1}^{2}\mathcal{A}_{j}=1~,
\end{equation}
wherein $\mathcal{R}$ is the reflectance, $\mathcal{A}_{1}$ the absorptance in the underlayer and $\mathcal{A}_{2}$ the absorptance in the overlayer (or generic city block) given by \cite{p95}:
\begin{equation}\label{4-280}
\mathcal{R}=\frac{\int_{0}^{\infty}\rho_{m}(\omega)\|A^{[0]+}(\omega)\|^{2}d\omega}{\int_{0}^{\infty}\|A^{[0]+}(\omega)\|^{2}d\omega}~~,~~
\mathcal{A}_{j}=\frac{\int_{0}^{\infty}\alpha_{2}^{[j]}(\omega)\|A^{[0]+}(\omega)\|^{2}d\omega}{\int_{0}^{\infty}\|A^{[0]+}(\omega)\|^{2}d\omega}~.
\end{equation}
We assume that $\|A^{[0]+}(\omega)\|^{2}$ is $\approx 0$ for $\omega>\omega_{max}$, so that $\mathcal{R}\approx \tilde{\mathcal{R}}$, $\tilde{A}_{j}\approx \tilde{\mathcal{A}}_{j}$, wherein
\begin{equation}\label{4-280}
\tilde{\mathcal{R}}=\frac{\int_{0}^{\omega_{max}}\rho_{m}(\omega)\|A^{[0]+}(\omega)\|^{2}d\omega}{\int_{0}^{\omega_{max}}\|A^{[0]+}(\omega)\|^{2}d\omega}~~,~~
\tilde{\mathcal{A}}_{j}=\frac{\int_{0}^{\omega_{max}}\alpha_{m}^{[j]}(\omega)\|A^{[0]+}(\omega)\|^{2}d\omega}{\int_{0}^{\omega_{max}}\|A^{[0]+}(\omega)\|^{2}d\omega}~.
\end{equation}
in which the various integrals can now be computed by any standard quadrature technique, the same being true of the integral over $z$ in $\alpha_{m}^{[j]}(\omega)$. To control the precision of the computation, we also compute the so-called normalized "output energy" $\mathcal{E}^{out}\approx \tilde{\mathcal{E}}^{out}=\tilde{\mathcal{R}}+\tilde{\mathcal{A}}_{1}+\tilde{\mathcal{A}}_{2}$ which, if the computation is exact, should be equal to the normalized "input energy" $\mathcal{E}^{in}=1$.
\clearpage
\newpage
\section{Numerical predictions of how variations of city density and city height affect the absorptance in the city/overlayer}
%
\subsection{Variation of city density for the case treated by Kham et al.}
\begin{figure}[ht]
\begin{center}
\includegraphics[width=11.5cm] {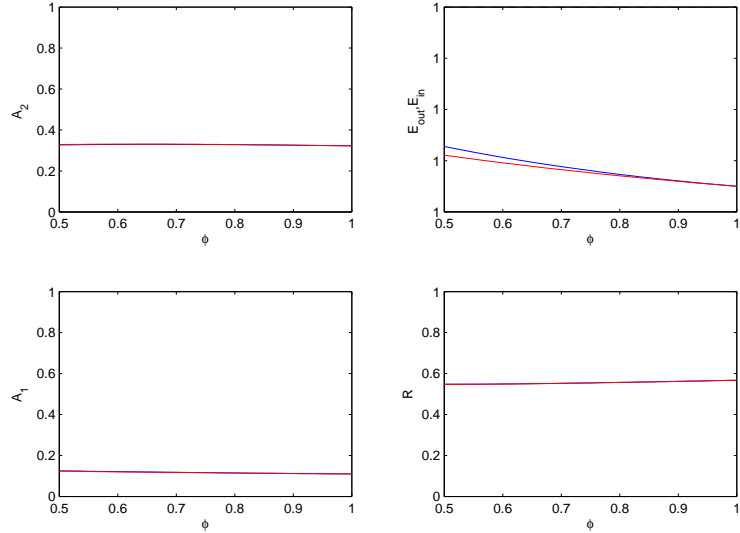}
 \caption{The upper left-hand panel depicts $\tilde{\mathcal{A}}_{2}$ versus $\phi$, the lower left panel $\tilde{\mathcal{A}}_{1}$ versus $\phi$, the lower right-hand panel $\tilde{\mathcal{R}}$ versus $\phi$ and the upper right-hand panel the normalized input (black line) and normalized output (red and blue curves) energy versus $\phi$. In all four panels, the blue curves apply to the periodic block model  of the city and the red curves to the corresponding overlayer model of the city. $h_{2}=30~m$. $w=10~m$, $\nu=2~Hz$..}
  \label{en-1}
  \end{center}
\end{figure}
We see in fig. \ref{en-1}, relative to $h_{2}=30~m$, that the overlayer and city absorptances are coincident and very slightly decreasing with $\phi$ which is almost the same behavior as that of the spectral absorptance and two transfer functions versus $\phi$ and thus supports, although weakly, the assertion in \cite{ks06} that increasing city density provokes a beneficial effect on the seismic response in the city. Of supplementary  interest is the fact that about a third of the incident energy is injected and spent in the city/overlayer, the rest being divided between absorbed energy in the underlayer and radiation damping, so that the sum of these three energies is equal to the incident energy.
\subsection{Variation of  city density for the case  of wider blocks than ones of the city treated by Kham et al.}
\begin{figure}[ht]
\begin{center}
\includegraphics[width=11.5cm] {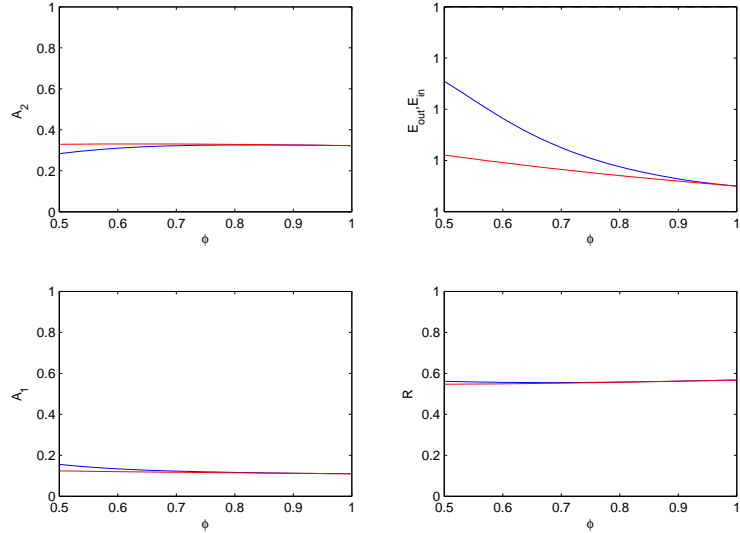}
 \caption{Same meaning of the panels as in fig. \ref{en-1}. In all panels, the blue curves apply to the periodic block model  of the city and the red curves to the corresponding overlayer model of the city. $h_{2}=30~m$,   $w=50~m$, $\nu=2~Hz$.}
  \label{en-2}
  \end{center}
\end{figure}
\begin{figure}[ptb]
\begin{center}
\includegraphics[width=11.5cm] {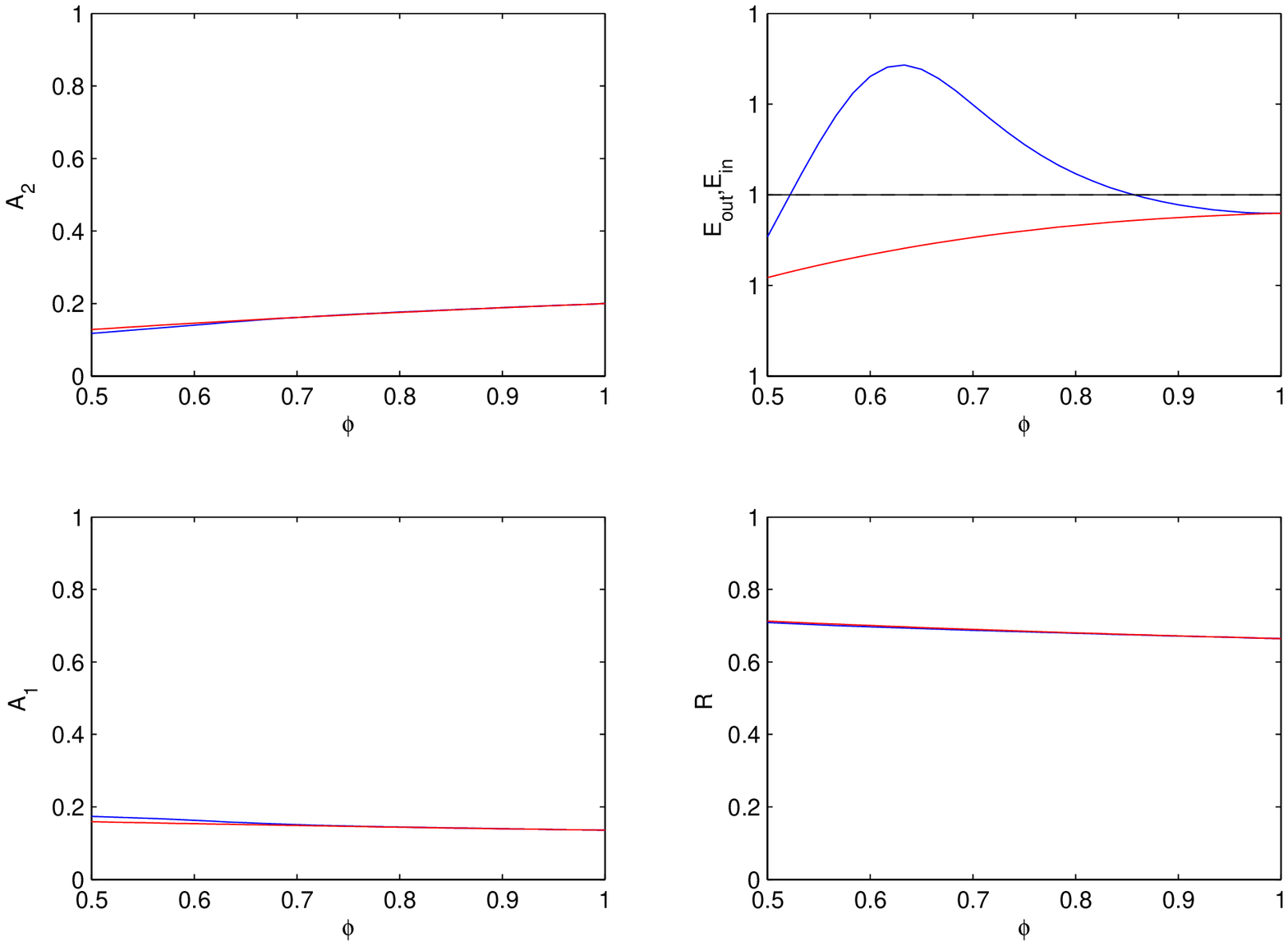}
 \caption{Same meaning of the panels as in fig. \ref{en-1}. In all panels, the blue curves apply to the periodic block model  of the city and the red curves to the corresponding overlayer model of the city. $h_{2}=60~m$,   $w=50~m$, $\nu=2~Hz$.}
  \label{en-3}
  \end{center}
\end{figure}
The wider block choice enables larger block heights to conform to more common aspect ratios. We  see in fig. \ref{en-2}, which again applies to $h_{2}=30~m$, that the overlayer and city absorptances are coincident, at least for the larger $\phi$ (because for smaller $\phi$, a larger portion of the motion occurs at the higher frequencies where the layer model becomes less apt to describe the response). Now the periodic block model predicts increasing city absorptance for smaller city densities and both models predict stable absorptance for larger city densities.

For the same block width but  a block height $h_{2}=60~m$, we observe in fig. \ref{en-3} that once again the layer model and periodic block model curves are nearly coincident, but now {\it the absorpance $A_{2}$ increases over the whole range of city densities}, which behavior was also previously observed in fig. \ref{att-02} relative to the spectral absorptance $\alpha_{2}$ at $2~Hz$.
\begin{figure}[ptb]
\begin{center}
\includegraphics[width=11.5cm] {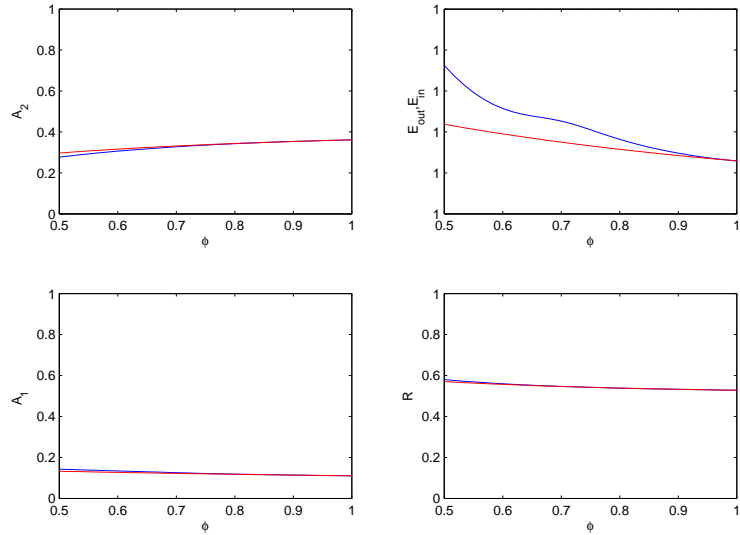}
 \caption{Same meaning of the panels as in fig. \ref{en-1}. In all panels, the blue curves apply to the periodic block model  of the city and the red curves to the corresponding overlayer model of the city. $h_{2}=90~m$,   $w=50~m$,  $\nu=2~Hz$.}
  \label{en-4}
  \end{center}
\end{figure}

In fig. \ref{en-4}, relative to $h_{2}=90~m$, the behavior of $\tilde{\mathcal{A}}_{2}$  is again increasing with $\phi$, but the amount of energy sent into the city is larger, attaining over a third of the incident energy for the densest city. Thus, from global (i.e.,  within the city) energy point of view (which is what counts in relation to damage), increasing the  city density does not  have the beneficial effect predicted in \cite{ks06,bg07} on the basis of transfer function and ground kinetic energy behavior.
\clearpage
\newpage
\subsection{Variation of  city height for the case  of wide blocks}
\begin{figure}[ht]
\begin{center}
\includegraphics[width=11.5cm] {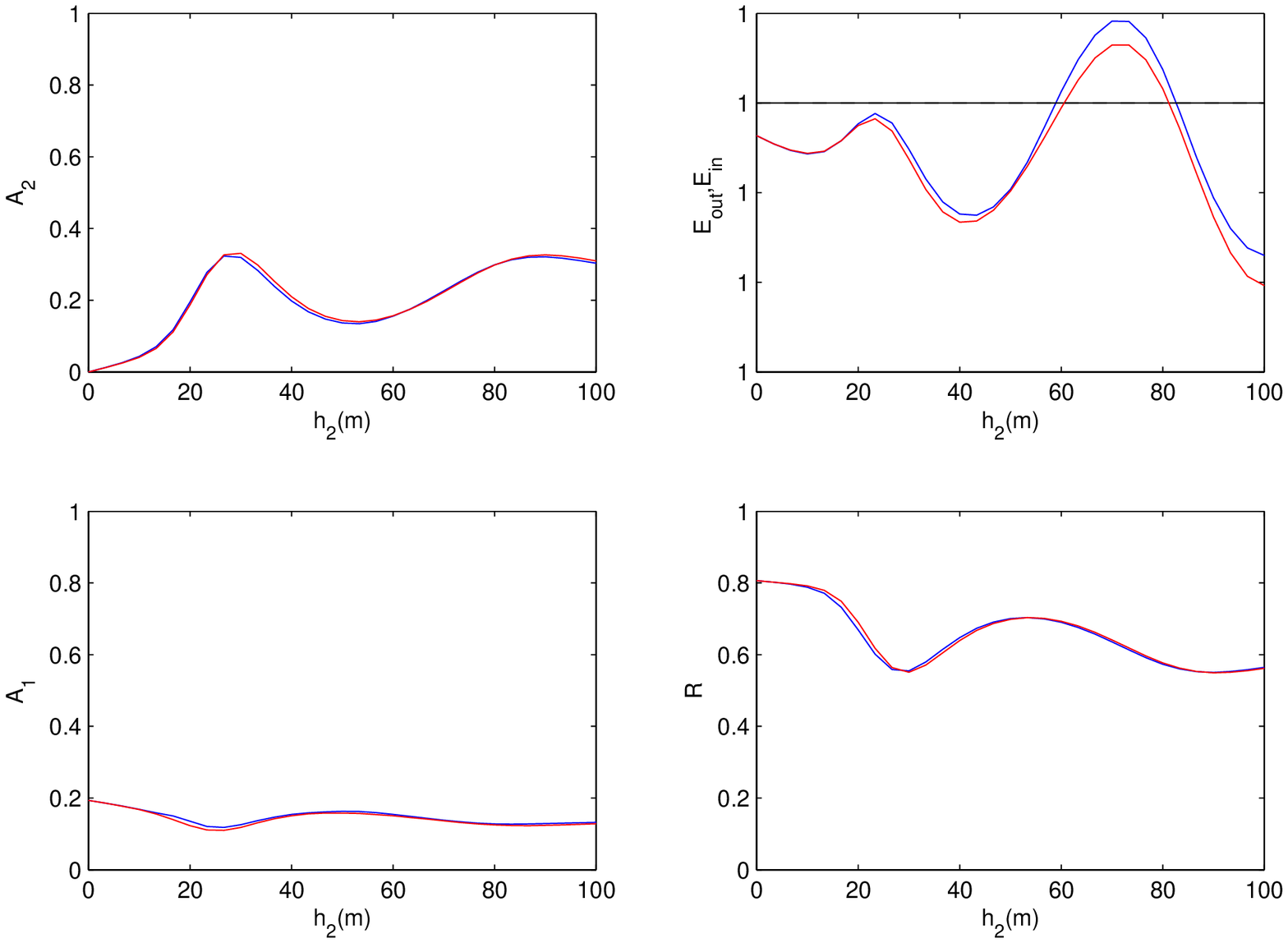}
 \caption{The upper left-hand panel depicts $\tilde{\mathcal{A}}_{2}$ versus $h_{2}$, the lower left panel $\tilde{\mathcal{A}}_{2}$ versus $h_{2}$, the lower right-hand panel $\tilde{\mathcal{R}}$ versus $h_{2}$ and the upper right-hand panel the normalized input(black line) and normalized output(red and blue curves) energy versus $h_{2}$. In all four panels, the blue curves apply to the periodic block model  of the city and the red curves to the corresponding overlayer model of the city.  $w=50~m$,  $d=75~m$ (i.e., $\phi=0.67$),    $\nu=2~Hz$.}
  \label{en-5}
  \end{center}
\end{figure}
Fig. \ref{en-5} concerns the effect of increasing block height (and/or overlayer thickness) $h_{2}$ on the way the incident energy is divided between the three regions of the configuration. It is observed that $\tilde{\mathcal{A}}_{2}$ is an oscillating function of $h_{2}$ with the long trend being, on the average, an increase of absorptance in the superstructure. This agrees with the behavior of the spectral absorptance as a function of $h_{2}$ at the characteristic frequency of the incident Ricker pulse, i.e., $f=\nu=2~Hz$, shown previously in fig.
\ref{hatt-01}. Again, one notes the large values of absorptance (up to $\sim 0.35$) when $A_{2}$ attains its relative maxima, but even for city heights as low as $20~m$ the absorptance is consequential. Thus, from the global energy point of view,  increasing the  city density doe not have a beneficial effect.
%
\section{Conclusion}
This investigation began with the simple observation that modern earthquake-prone cities (often built on rather soft soil) are growing in height and density (as well as population) with time. A natural question  is then:  can this trend have a significant impact on the risk of damage  (and casualties to the populations \cite{b99}) during seismic events?

 The majority opinion in response to this question, based on quite elaborate numerical and experimental studies (see, e.g., \cite{ks06,sb16}), is that increasing city density will certainly have an effect on the seismic vulnerability and that this effect will be 'beneficial'.  This raises the second question of how these 'beneficial' effects are measured.

  In our investigation, we  focused our attention on a single article, that of Kham et al. \cite{ks06}, dealing with these questions. These authors share the majority opinion and provide the following answer to the second question:  the beneficial effect manifests itself essentially in the measurements associated with displacement transfer functions (TF's) on the ground and top of the city and cumulative (over the signal duration) kinetic energy (KE) at one or two points on the ground.

  This raises a third question: are the one or two measurements of TF's and ground KE informative enough to decide whether the buildings in a city will shake more (adverse effect) or less (beneficial effect) violently? It is not easy to answer this question without being influenced by the thought that, of course, the city being a quite heterogeneous structure, would require, to describe its motion in detail, a large number of seismometers to be placed (and synchronized) at various locations of the ground and within the buildings and that this is economically not feasible. But what is not feasible in the field and perhaps in a laboratory experiment is feasible, thanks to modern computers,  whereby the motion at any point of  structures, even those as complex as cities, can be computed \cite{gtw05,kj06,t10,it14,gc16}. But  this is a formidable undertaking due to the many types of cities, districts therein, and buildings which means that a very large number of configurational parameters will have to be varied to get a decent picture of how they influence the amount of shaking in the city. Fortunately, mathematical theory can be a powerful ally and enable predictions, or at least serve as a guide to those who prefer numerical experimentation, of how a city responds to a seismic wave.

  After having opted for a theoretical approach, the next question is: what type of city configuration should be studied and is it representative of the majority of modern cities? The answer is easy to obtain because there do not exist all that many city configurations which can be studied theoretically. If we exclude the case of older towns or 'cities' composed of several isolated buildings (e.g., \cite{gtw05,t10}), then we know of only three types of such cities: a random city \cite{ca01}, a periodic city (e.g., \cite{g05,ks06,gw08}), and a homogenized city {e.g.,\cite{br04}). We are not of the opinion that modern cities have layouts which are random in nature, so that a theoretical undertaking such as ours  must deal with either (or both) the periodic or homogenized cities. There are plenty of arguments in favor of the periodic city model (especially for modern cities) although it is, of course, an idealization, like any object accessible to a decent theoretical analysis. However, since the subject of seismic response in periodic cities has been relatively-often belabored, we chose herein to employ the rigorous results that such a model enables only as a reference by which to judge the quality of the approximate theoretical and numerical results we derived from a homogenized-city model.

  Homogenization is a very old, and certainly useful, device for getting a theoretical grasp on complicated geophysical problems. For instance, an object such as the earth is often thought of as being composed of a superposition of distinct, homogeneous layers \cite{ejp57,k64} and from this emerge the notions of compressional and shear body  and surface waves that account for easily-recognizable features in real seismograms. With this in mind, we decided to employ a (linear, homogeneous, isotropic) layer model of the city in order to provide theoretical and numerical answers to the aforementioned questions. Our layer model is the outcome of an approximation procedure that will be published elsewhere whereby a periodic block model of the city is shown to produce the same response to an incident seismic body wave as a  homogeneous layer whose thickness equals the height of the blocks and shear modulus is the product of the block shear modulus with the city density. Naturally, this equivalence holds only approximately, and under certain conditions: (i) the sources of the seismic solicitation are far (and underneath) from the ground on which the city rests so that the incident wave can be considered to be a SH body wave, (ii) the city is of the 2D variety such that the scattered waves are also and uniquely SH, (iii) the frequency is so low that only two body waves have significant amplitudes in the city as well as in the soil layer and hard half infinite underground, (iv) the city is dense, which means that the blocks (or buildings) occupy, at their base, a large fraction of the total area of the city. Since the cities we study are relatively-dense and the seismic phenomena we deal with are of the low frequency variety (of the order of $1~Hz$) it is not absurd, a priori, to employ such a layer model for the problem at hand, provided, of course, that we can evaluate (as we have done) the accuracy of the theoretical predictions to which this model leads by comparison with rigorous reference predictions.

  Although our approach is geophysical, the problem we address is not typically geophysical in that the latter often deals with trying to explain spectral features of seismograms in order to gain information on the seismic sources, as well as the composition and geometry of a geophysical object (such as the earth) whereas here we start from  supposedly-known sources and object (the city and the site on which it rests) and want to find out how violently it reacts to a seismic solicitation. More specifically, we wanted to find out if certain types of measurements, which are more of the strength-of-motion than spectral variety, can  inform us sufficiently for it to be possible to decide scientifically whether or not increasing city density, and other to-be-identified key parameters that increase with time, produce a beneficial effect or not for the residents  of the buildings in a modern city.

  To briefly resume what we have found, probably the most important points are:\\
  (a) potential damage in a city cannot be predicted uniquely on the basis of whether the modulus of  transfer functions at a few points on the boundary of a generic block or building of the city increases or decreases,\\
  (b) due to the heterogeneity of response in the buildings or blocks of the city, the motion throughout the latter must be measured and combined into a single entity (called 'overlayer flux' which, at each frequency, obeys, together with the underlayer, half-space and incident fluxes, a conservation law,\\
  (c) even the flux concept is not sufficient to describe the global response of the city since it takes no account of the characteristics of the incident seismic pulse, i.e., its duration or spectral features, so that the flux must be multiplied by the pulse spectrum and integrated over all frequencies to obtain a measure of the energy injected into (the blocks or buildings of) the city, this energy being the true measure of potential damage (other factors, such as the particularities of building conception and construction, contribute of course to the degree of damage, but are either not accounted for, or integrated in variations of the homogenized constitutive  properties of the buildings, blocks or overlayer),\\
  (d) measurements of the global energy injected into the buildings, blocks or their overlayer-like surrogate reveal that this energy does not always decrease with increasing city density as it should if the effect were 'beneficial', but, on the contrary, increases with city density for certain building heights,\\
  (e) the long-scale trend of increasing city height is to increase the amount of energy sent into the city,\\
  (f)  the energy sent into the built component of a modern city can attain more than a third of the incident seismic energy, this constituting the most frightening discovery of our investigation.\\

Most of these predictions result from the layer model of the city and have been verified by means of the periodic block model. The next step should be to employ the layer model and bring the seismic sources closer to the ground so as to be able to excite surface waves \cite{gw05a,gw05b} which surely will give rise to responses with higher quality factors and consequently longer durations as observed in sites such as Mexico City \cite{gbc02}. An interesting task would also be to generalize the layer model for P and SV waves so as to be able to more completely predict the seismic motion in the city.

\end{document}